\PassOptionsToPackage{prologue,dvipsnames}{xcolor}

\documentclass[sigplan,nonacm,10pt]{acmart}

\usepackage{tcolorbox}
\usepackage{ebproof}
\usepackage{subcaption}
\usepackage{listings}
\usepackage{wrapfig}
\usepackage{multirow}
\usepackage{wasysym}
\usepackage{syntax}
\usepackage{tikz}
\usetikzlibrary{decorations.pathreplacing,angles,quotes}
\usepackage{adjustbox}
\usepackage{comment}
\usepackage{tcolorbox}
\usepackage{svg}
\usepackage{macrodefs}
\usepackage[inline]{enumitem}
\usepackage{pifont}

\definecolor{vgreen}{RGB}{104,180,104}
\definecolor{vblue}{RGB}{49,49,255}
\definecolor{vorange}{RGB}{255,143,102}

\makeatletter
\DeclareRobustCommand*{\escapeus}[1]{%
    \begingroup\@activeus\scantokens{#1 }\endgroup}
\begingroup\lccode`\~=`\_\relax
    \lowercase{\endgroup\def\@activeus{\catcode`\_=\active \let~\_}}
\makeatother

\setcopyright{none}

\theoremstyle{definition}
\newtheorem{theorem}{Theorem}[section]
\newtheorem*{theorem*}{Theorem}

\newtheorem{lemma}[theorem]{Lemma}

\newtheorem{definition}[theorem]{Definition}

\renewcommand\footnotetextcopyrightpermission[1]{}
\begin{document}

\title{\codename{}: A General-Purpose Timing-Safe Hardware Description Language}
\pagestyle{plain}

\newcommand{\nusaffiliation}{
\affiliation{%
  \institution{National University of Singapore}
  \country{Singapore}
}
}

\author{Jason Zhijingcheng Yu}
\authornote{Equal contribution.}
\email{yu.zhi@comp.nus.edu.sg}
\orcid{0000-0001-6013-157X}
\nusaffiliation{}

\author{Aditya Ranjan Jha}
\authornotemark[1]
\email{arjha@comp.nus.edu.sg}
\nusaffiliation{}

\author{Umang Mathur}
\email{umathur@comp.nus.edu.sg}
\nusaffiliation{}

\author{Trevor E. Carlson}
\email{tcarlson@comp.nus.edu.sg}
\nusaffiliation{}

\author{Prateek Saxena}
\email{prateeks@comp.nus.edu.sg}
\nusaffiliation{}

\begin{abstract}
Expressing hardware designs using hardware description languages (HDLs) routinely involves using stateless signals whose values change according to their underlying registers.
Unintended behaviours can arise
when the stored values in these underlying registers are mutated while
their dependent signals are expected to remain constant across
multiple cycles.
Such \emph{timing hazards} are common because, with a few exceptions, existing HDLs lack abstractions for values that remain unchanged
over multiple clock cycles, delegating this responsibility to hardware designers.
Designers must then carefully decide whether a value should remain unchanged,
sometimes even across hardware modules.
This paper proposes \codename{},
an HDL which statically prevents timing hazards with a novel
type system.
\codename{} is the only HDL we know of that guarantees
\emph{timing safety}, i.e., absence of timing hazards, without
sacrificing expressiveness
for cycle-level timing control or dynamic timing behaviours.
Unlike many HLS languages that abstract away the differences between registers and signals, \codename{}'s
type system exposes them fully while capturing the timing relationships between
register value mutations and signal usages to enforce timing safety.
This, in turn, enables safe composition of communicating hardware modules by static
enforcement of \emph{timing contracts}
that encode timing constraints on shared signals.
Such timing contracts can be specified parametric on
abstract time points that can vary during run-time, allowing the type system to statically express
dynamic timing behaviour.
We have implemented \codename{} and successfully used it to implement key timing-sensitive modules,
comparing them against open-source SystemVerilog counterparts to demonstrate the practicality and expressiveness of the generated hardware.
\end{abstract}

\maketitle

\section{Introduction}
\label{sec:intro}

Hardware description languages (HDLs) shape the way people
think about and describe hardware designs.
Ideally, an HDL should provide easy-to-use abstractions
for hardware designers to
express their intention precisely and correctly.
The concurrent and continuous behaviour of hardware makes this goal challenging to achieve.

Unlike software programs, where values are all persistent (stored either in registers or in memory),
hardware designs involve separate notions of \emph{signals} and \emph{registers}.
While a register can store persistent values and be assigned new values
every cycle, signals are stateless, with their values changing with the registers
they depend on.
{
If the hardware designer expects a signal to remain unchanged
across multiple cycles, they must explicitly ensure the stored values of their
underlying registers do not change.
The incorrect timing of register mutation
(i.e., change of the stored value in a register) and signal use thus easily introduces
invalid or wrong values during run-time, and may even expose the hardware
design to time-of-check-to-time-of-use (TOCTOU) attacks.%
}
We call such problems \emph{timing hazards}.
The problem of timing hazards is further exacerbated
by the concurrent nature of hardware designs:
A hardware design commonly consists of
large numbers of modules which are executing in parallel
and communicating with each other via shared signals.

Most existing HDLs such as SystemVerilog~\cite{18002017IEEEStandard2018}, VHDL~\cite{VHDLRef},
and Chisel~\cite{bachrachChiselConstructingHardware2012}
do not catch timing hazards at compile time,
leaving designers to discover these issues only during simulation.
Designers frequently seek help on discussion forums simply to pinpoint the origins of the errors~\cite{chunduri1011OverlappingMealy2020,titanSynchronizingMultiplierAdder2014,aluissue}.
Timing hazards are prevalent even among experienced designers
and in widely used open-source hardware components, as is shown by several real-world examples given in Appendix~\ref{subsec:error}.

{A principled way to eliminate timing hazards is to forbid them in the HDL itself.
We call such an HDL that only allows designs without timing hazards \emph{timing-safe}.}
The key challenge to achieve timing safety while also providing enough expressiveness for writing general-purpose hardware designs.
{Some existing HDLs can provide timing safety but only at a significant cost of expressiveness,
making
them only suitable for specific applications.
For example,
high-level synthesis (HLS) languages~\cite{xls,systemc,baayDigitalCircuitsClaSH2015}
offer software-like programming models for hardware design.
In these languages,
timing hazards are not a concern because they abstract away both cycle latencies and the distinction between wires and registers, effectively treating all values as persistent, similar to variables in software programming.}
This expressiveness for cycle-level control and wires is, unfortunately, absent in such languages. This is
an essential abstraction in general-purpose hardware designs, especially where performance is a priority.
Consequently, the applicability of HLS languages is commonly
limited to speeding up algorithms with programmable hardware (e.g., FPGA).
Other {timing-safe} languages focus only on specific types of hardware designs,
such as CPU stages~\cite{zagieboyloPDLHighlevelHardware2022}
and static pipelines~\cite{nigamModularHardwareDesign2023,spade}.

We present \codename{},
the first HDL we know of that \textbf{guarantees \emph{timing safety} while maintaining
expressiveness for general-purpose hardware design use cases}.
{\codename{} is general-purpose in the sense that the designer retains full
control of the cycle-level timing and register states in RTL, unlike HLS languages, and is
not limited to design use cases such as CPU-level abstractions and static pipelines.
In particular,}
it allows hardware designers to seamlessly specify cycle-level delays and
to express whether a value is stored in a register.
It also supports expressing
hardware designs with dynamic timing behaviours easily.

\codename{} achieves timing safety statically
with a novel type system which
captures the timing relationships between register mutations and
use of signals.
It performs type checking that
reasons about whether each use of signal takes place in a time window
throughout which it carries an unchanging and meaningful value,
and rejects code that is not timing-safe.
Designs written in \codename{} can thus specify precise cycle-level behaviour and register updates.
This is in contrast to HLS languages~\cite{xls} that hide wires and cycles beneath
their abstractions.
Across hardware modules, \codename{}'s type system
guarantees safe composition by statically checking against
\emph{timing contracts}, which specify constraints regarding
communicated signals, including constraints about when such signals must be kept unchanged.
Although \codename{}'s type checking is entirely static,
it explicitly allows dynamic timing behaviours, i.e.,
the number of cycles for a behaviour of the hardware design can
vary during run-time (e.g., caches).
The type system achieves this by capturing time not in terms of an absolute (fixed) number of cycles, but instead as abstract time points that correspond
to events that may occur arbitrarily late, for example, the event corresponding to the receipt of data
from another module.
This is in sharp contrast to recent work~\cite{nigamModularHardwareDesign2023}
in which the proposed type system
only allows expressing designs with fixed static timing behaviours.

We have implemented \codename{} (Section~\ref{sec:impl}).
The \codename{} compiler performs type checking and
compiles \codename{} code to SystemVerilog.
Our evaluations highlight
the expressiveness and practicality of \codename{}
(Section~\ref{sec:eval}).
Designs written in \codename{} can be integrated in
existing code bases in other HDLs, thus allowing incremental
adoption and making \codename{} immediately useful.
We have successfully used \codename{} to implement a diverse set of
10~latency-sensitive components ranging from an AES accelerator~\cite{opentitanaes}
to
a page table walker in a \riscv{} CPU~\cite{zarubaCostApplicationclassProcessing2019}.
Despite the \codename{} compiler being an
early-stage prototype,
when compared with open-source SystemVerilog implementations,
the \codename{} implementations show practical overhead
averaging $4.50\%$ for area and $3.75\%$ for power.
We have made \codename{} open-source for public use
at \url{https://github.com/jasonyu1996/anvil}.

\nparagraph{Our Contributions}{%
We introduce \codename{}, an HDL with a novel type system
that guarantees timing safety without sacrificing
expressiveness, e.g., for cycle-level control and dynamic timing behaviours.
\codename{} allows for general-purpose hardware design use cases and
integration with existing SystemVerilog projects.
}

\begin{figure}[t]
    \centering
    \begin{minipage}{.31\linewidth}
    \begin{codebox}
    \begin{lstlisting}[language=verilog]
module Memory (
 ...
 input [7:0] inp,
 input req,
 output [7:0] out
);
\end{lstlisting}
    \end{codebox}
    \end{minipage}
    \hfill
    \begin{minipage}{0.66\linewidth}
    \includegraphics[width=\linewidth]{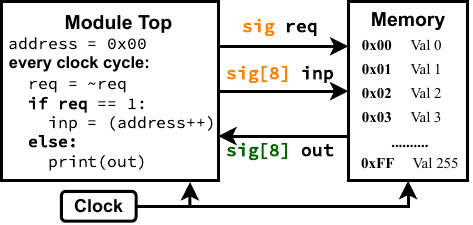}
    \end{minipage} \\
    \begin{tikzpicture}
        \node [anchor=south west,inner sep=0] (image) at (0,0)
        {\resizebox{\linewidth}{!}{\includesvg{images/egSec2merged.svg}}};

        \draw[decoration={brace,mirror,raise=2pt},decorate]
            (6.74,0.75) -- node[below=3pt] {\tiny[T, T+2)} (7.72,0.75);
        \draw[decoration={brace,mirror,raise=8pt},decorate]
            (7.74,0.75) -- node[below=9pt] {\tiny[T+2, T+3)} (8.25,0.75);
    \end{tikzpicture}
    \caption{Module \texttt{Top} interfaced with \texttt{Memory}.}
    \label{figure:cbeg}
\end{figure}

\section{Motivation}

\label{sec:bg}

{%
The motivation of our work stems from the susceptibility of RTL designs to
timing hazards due to limitations of \textit{de facto} standard HDL abstractions.%
}

\subsection{Example of a Timing Hazard}

Consider the interface of a memory module in SystemVerilog in
Figure~\ref{figure:cbeg} top left.
Unlike software,
hardware modules communicate using signals that can be continuously read and updated.
Consider an interfacing hardware module (Figure~\ref{figure:cbeg}, top right), \texttt{\modulename},
which reads a value from a memory module with the same interface. The implementation of \texttt{\modulename} sends an address as a request and expects to read the output in the following cycle.
However, the circuit outputs are incorrect, as evident
when the system is simulated (Figure~\ref{figure:cbeg}, bottom left).
The culprit is an unexpected timing delay.
The module \texttt{\modulename} is written under the assumption that the memory subsystem responds precisely one clock cycle after the \texttt{req} signal is set. However, the memory subsystem takes two cycles to process the lookup request and return the output.

\begin{figure*}[t]
    \centering
    \includegraphics[width=\textwidth]{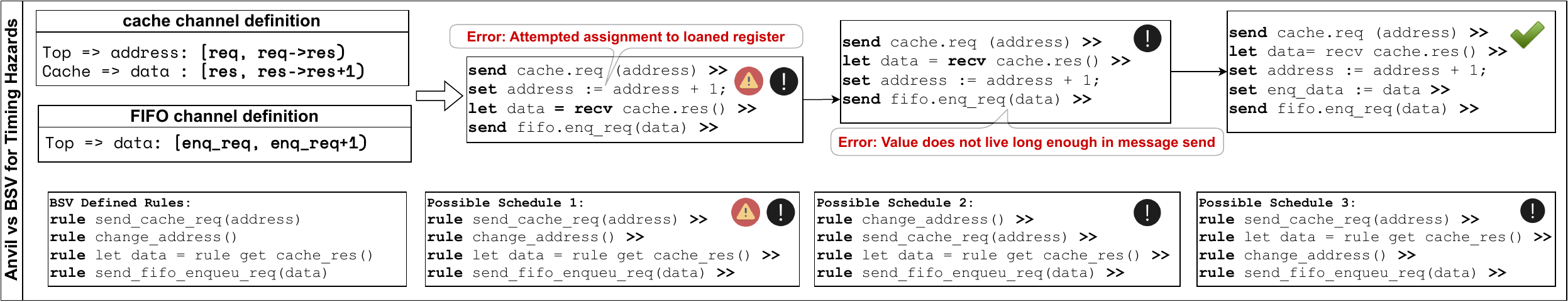}
    \caption{Top: Anvil guiding designer through timing safe design. Bottom: BSV timing unsafe schedules.}
    \label{fig:bsvfail}
\end{figure*}

In more detail, the module \texttt{\modulename} requests address \texttt{0x00} by setting the \texttt{req} signal high during cycle $[0,1)$.
It expects the output in the next cycle, but the memory has not finished dereferencing the input \texttt{address}. The memory stops processing since the \texttt{req} signal is unset in $[1,2)$. When \texttt{req} is set again in $[2,3)$ with address \texttt{0x01}, the memory is still resolving \texttt{0x00}, returning \texttt{Val 0} in $[3,4)$.
Meanwhile, the input address changes from \texttt{0x01} to \texttt{0x02}. When \texttt{req} is set again in $[4,5)$, the memory starts processing \texttt{0x02}, skipping \texttt{0x01}. As a result, unexpected outputs are observed, and only half of the requested addresses are dereferenced.

The above example illustrates a classic case of a timing hazard, where unintended values are used or values in use are changed unexpectedly.
Here, the module \texttt{\modulename} modifies its input while the memory still processes the address lookup request. It also reads the output before it is ready.

\subsection{Timing Hazards in Existing HDLs}
\label{subsec:existing-hdls}

Timing hazards arise in SystemVerilog and VHDL,
two standard and most widely used HDLs,
as they lack an abstraction for
the designer to express values that are sustained across multiple cycles.
These languages also do not provide a mechanism to encode timing constraints pertaining to
register assignments and use of signals shared
between communicating modules.
{%
The abstraction that SystemVerilog and VHDL provide over registers and signals specifies their relationships within a single, non-specific cycle. The designer defines how each register is updated based on the existing register state, and signal values are then automatically updated accordingly.
In other words, signals are essentially pure functions of the current register state; when they are referenced in the code, they simply carry the values of the current moment.
Such an abstraction makes it difficult to express intended relationships between signal values across multiple cycles.
For example, a SystemVerilog implementation of the \texttt{\modulename} module in Figure~\ref{figure:cbeg} does not convey the intent that \texttt{req} and \texttt{inp} should remain steady for two consecutive cycles, or that \texttt{out} should be meaningful only in the following cycle.
Other HDLs—including many newer ones that aim to raise the abstraction level for hardware design (e.g., Chisel~\cite{bachrachChiselConstructingHardware2012} and SpinalHDL~\cite{spinalhdl})—follow the same fundamental paradigm for describing RTL designs as SystemVerilog and VHDL, and are therefore similarly susceptible to timing hazards.%
}

{
Some popular HDLs provide different abstractions than
SystemVerilog and VHDL but are still unable to
avoid timing hazards.}
Bluespec SystemVerilog (BSV)~\cite{BluespecSystemVerilogReference2008}, for example,
provides the abstractions of rules and methods.
Rules are bundled hardware behaviours that execute atomically.
Modules communicate through invoking each other's exposed methods, which add to the behaviours to
be executed.
The BSV compiler generates hardware logic to
choose rules to execute in each cycle.
For example, consider Figure~\ref{fig:bsvfail}.
If \texttt{\modulename} reads a value from a cache and enqueues it into a FIFO queue that only accepts requests when it is not full, the design would typically use two rules:
one to invoke the read method of the cache, and another to enqueue the retrieved value into the FIFO.
BSV's scheduler ensures that, in each cycle, rules that execute do not conflict (i.e., they do not mutate the same registers), and each rule executes atomically.
However, rules only specify operations for the \emph{current} cycle, and scheduling is performed independently for each cycle.
BSV does not reason about behaviours that span multiple cycles~\cite{BluespecSystemVerilogReference2008}.

In the example, if the module \prog{Top} retrieves a value from a cache and sends it to a FIFO, \codename{} enforces the timing contract by detecting violations and guiding the designer toward a timing-safe implementation, as shown in Figure~\ref{fig:bsvfail} (top).
BSV, on the other hand, may still generate a conflict-free schedule that is \emph{timing-unsafe} because it does not capture inter-cycle constraints in its scheduling model.

{%
\nparagraph{Root Cause: HDL Abstractions}{%
In summary, the root cause behind the susceptibility of many
popular HDLs to timing hazards lies in the abstractions
they provide.
In particular, their abstractions do \emph{not}
express \emph{the designer's intent} concerning
when a signal is expected to carry
a meaningful value and in which time window
the value is expected to remain steady.
As such, we provide a novel solution which in  new HDL design
rather than basing it on existing ones.%
}}

{
\subsection{Need for Timing-Safe HDL Abstractions}
\label{subsec:hdl-abstractions}

In this paper, we tackle the problem of timing hazards by creating
\emph{timing-safe HDL abstractions} to capture the designer's intent
regarding register and signal uses across cycles and in turn
prevent timing hazards.
An alternative approach is to apply verification techniques to
designs expressed in existing HDLs~\cite{moszkowskiTemporalLogicMultilevel1985,guptaFormalHardwareVerification,kernFormalVerificationHardware1999,witharanaSurveyAssertionbasedHardware2022}.
Such techniques attempt to verify that certain properties
about a design (e.g., user-specified SystemVerilog assertions) hold,
either statically through formal verifications (e.g.,
model checking with Cadence JasperGold~\cite{JasperFormalVerification} or
Yosys SMT-BMC~\cite{YosysHQYosys2025}) or
dynamically through testing (e.g., simulation-based verification
with UVM~\cite{IEEEStandardUniversal2020} or cocotb~\cite{Cocotb}).
This approach is general and may easily extend to other
properties about an RTL design beyond timing safety.
It is also readily applicable to existing code bases and
requires no switching to a new language by the designer.

\begin{figure}[t]
    \centering
    \includegraphics[width=\linewidth]{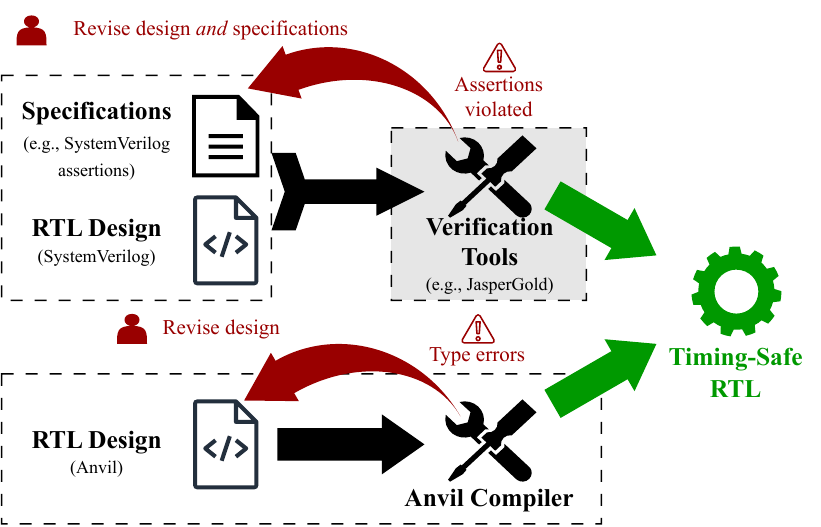}
    \caption{{High-level comparison between the flows enabled by
    verification- (top) and language-based (bottom) approaches.
    Steps involving manual effort are marked with the \emph{person} icon.
    White and gray dashed boxes represent design and verification stages,
    respectively.}}
    \label{fig:verification-comparison}
\end{figure}

However, we have been motivated to focus on a \emph{language-based} approach because of its unique advantages.
As illustrated in Figure~\ref{fig:verification-comparison}, a language-based approach can \emph{preclude} designs with timing hazards \emph{during} development. In contrast, verification detects timing hazards only \emph{after the fact}, in a separate verification stage. This allows a faster and more integrated feedback loop.
Through a language-based approach, the language abstractions themselves directly express the properties to be checked, for example,
as part of a type system.
A verification-based approach, on the other hand, requires manually specified, implementation-specific assertions to fill in missing information in the HDL abstraction.
These assertions are error-prone and a burden to maintain.
A language-based approach can also present a more abstract model for reasoning about timing hazards efficiently.
This avoids the state explosion problem with verification~\cite{clarkeModelCheckingState2012}.
For example, bounded model checking may fail to report a violation even at large depths because of the prohibitive size of the model generated from SystemVerilog code.
In Appendix~\ref{appx:verification}, we present a concrete example comparing \codename{}
---the language-based solution proposed in this paper---with verification-based methods to illustrate these points further.%
}

\subsection{Goal: a Timing-Safe and Expressive HDL}
\label{subsec:verification}

{Some existing HDLs do provide timing safety.}
{However, they}
face challenges in maintaining expressiveness.
Some high-level synthesis (HLS) languages~\cite{xls} provide
abstractions of persistent values similar to variables in software programs.
They abstract away certain aspects of hardware design such as register placements
and cycle latencies.
While their abstractions directly prevent timing hazards, they lack the precise
timing and register control desired in general-purpose hardware design use cases,
especially when the design needs to be latency-sensitive or efficient.

The closest prior work to ours is the Filament HDL~\cite{nigamModularHardwareDesign2023}.
Filament exposes cycle latencies and
registers to the designer, and prevents timing hazards through its type system centred
around \emph{timeline types}.
A timeline type encodes constraints regarding the time window in which each signal
carries an unchanging value
that can be used. Timeline types also serve to define contracts at module interfaces, allowing for safe composition of modules.
Our example memory module can be augmented with such a contract
which requires \texttt{input} and \texttt{req} to remain constant during $[T, T+2)$,
and the output to remain constant in $[T+2, T+3)$.
Figure~\ref{figure:cbeg} (bottom, right)
illustrates the output waveform for a system using this contract.
However, the timeline type and the contract it represents only capture
timing intervals whose duration is fixed to be a statically determined, constant number of cycles.
Correspondingly, Filament only aims to support pipelined designs with static timing.
This prevents Filament from expressing
common hardware designs such as
caches and page table walkers that
exhibit dynamic timing behaviour.

To see why this is the case,
consider a memory subsystem with a cache. Its timing behaviour varies significantly between a cache hit and a cache miss. If the designer chooses a conservative upper bound statically
on the response time to accommodate both cases, the static timing contract would prevent \issuename{}s but nullify the advantage of caching. Figure~\ref{fig:cacheWave} (left) illustrates the output waveform for such a system, where the contract uses the worst-case delay. In such cases, one must trade off the flexibility of
dynamic latencies for the static guarantee of timing safety.

\begin{figure}[t]
    \centering
    \begin{tikzpicture}
        \node [anchor=south west,inner sep=0] (image) at (0,0)
        {\resizebox{\linewidth}{!}{\includesvg{images/mergedCacheNew.svg}}};

        \draw[decoration={brace,mirror,raise=2pt},decorate]
            (0.9,0.4) -- node[below=3pt] {\tiny[T, T+3)} (2.24,0.4);
        \draw[decoration={brace,mirror,raise=2pt},decorate]
            (2.25,0.4) -- node[below=3pt] {\tiny[T+3, T+4)} (2.7,0.4);
    \end{tikzpicture}
    \caption{Cache output waveform expressed safely with static (left) and dynamic (right) timing contract.}
    \label{fig:cacheWave}
\end{figure}

\begin{figure*}[t]
    \centering
    \includegraphics[width=\textwidth]{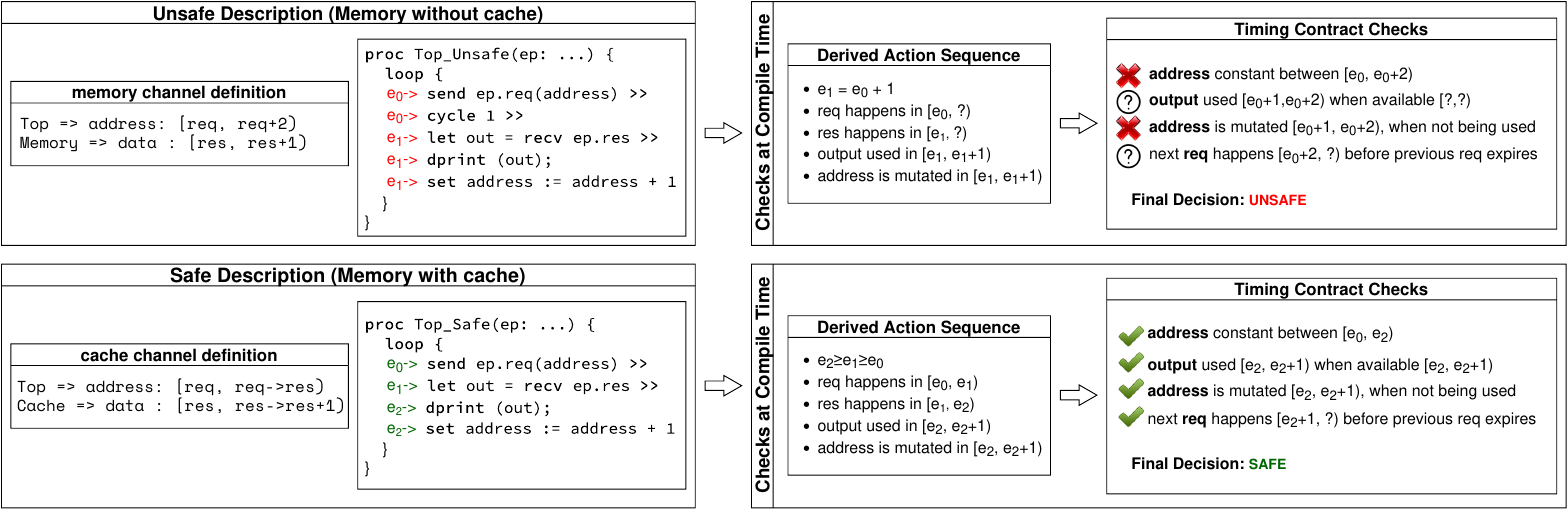}
    \caption{Anvil checking the unsafe version of \texttt{\modulename} interfacing with memory subsystem with and without cache.}
    \label{fig:anvilChecker}
\end{figure*}

\section{Timing Safety with \codename{}}

We present \codename{}, an HDL with a novel type system
that statically guarantees timing safety while retaining
the level of expressiveness required for a general-purpose
HDL.
Unlike HLS languages that abstract away registers and cycle latencies,
\codename{} gives the designer full control over register mutations and cycle latencies.
And unlike Filament~\cite{nigamModularHardwareDesign2023}, \codename{}'s type system
can capture and reason about timing that varies during run-time.
\codename{} is thus
able to enforce \emph{dynamic timing contracts} across modules and
precisely express hardware designs with dynamic timing behaviours.

\nparagraph{Channels}{%
\codename{} models hardware modules as communicating processes~\cite{hoareCommunicatingSequentialProcesses2000}.
It allows specifying modules with a process abstraction, using the keyword \prog{proc}.
A pair of communicating processes can share a bidirectional \emph{channel},
through which they send and receive values.
Channels are stateless and both sending and receiving are blocking.
Channels are the only way for processes to communicate.%
}

\nparagraph{Events}{%
A central concept that enables \codename{} to reason about
dynamic timing is \emph{events}.
Events are abstractions of time which may or may not statically map to a fixed cycle.
The start of every clock cycle is an event that is statically known (constant).
An example of a dynamic event is when two processes exchange a value through the channel.
As described above, sending and receiving values on a channel are blocking.
The exchange of the value thus completes at a time both sides agree on:
when the sender signals the value is valid and the receiver acknowledges.
The completion of this value exchange defines a dynamic event that may correspond to
varying clock cycles during run-time.
{Note that both events and channel-based communication are only
abstractions that \codename{} provides, and under the hood,
do not imply overhead in the resulting RTL design (see Section~\ref{sec:impl}).}%
}

\nparagraph{Event Graphs}{%
A key observation enables \codename{} to reason about events: even though
we cannot statically know which exact cycle an event may correspond to,
we know of the relationships among events.
For example, we can statically obtain that event $e_1$ corresponds to exactly two cycles
after the cycle $e_2$ corresponds to, and event $e_3$ corresponds to
the first time a specific value is exchanged on a channel after the cycle $e_2$ corresponds
to.
Such relationships form an \emph{event graph} (Section~\ref{subsec:event-graph}) which
serves as the basis for \codename{}'s type system (Section~\ref{safetychecks} and Appendix~\ref{appx:details}).
}

\nparagraph{Lifetimes and Dynamic Timing Contracts}{%
\codename{}'s type system uses events to encode
the lifetime of a value carried by a signal.
The lifetime of a value is identified by a start and an end event, between which the value is expected to remain steady.
Channel definitions in \codename{} specify the timing contracts for the exchanged values.
Since events can be bound to varying concrete clock cycles at runtime,
such timing contracts can capture \emph{dynamic}
timing characteristics.
Enforcement of timing contracts ensures timing-safe composition of two processes
when the events mentioned in the timing contract are known to both processes, e.g.,
when they correspond to value exchanges on the same shared channel.

}

\nparagraph{Example: \codename{} in Action}{%
Figure~\ref{fig:anvilChecker} illustrates how \codename{}'s type system distinguishes between safe and unsafe process descriptions.
The description \prog{proc Top_Unsafe} is \codename{}'s representation (simplified for understanding) of the same circuit \prog{Top} shown in Figure~\ref{figure:cbeg}.
In contrast, \prog{proc Top_Safe} captures the timing characteristics of the memory subsystem with a cache, as depicted in Figure~\ref{fig:cacheWave} (right).
\codename{} first derives the action sequence and then verifies whether the process description adheres to the constraints specified by the timing contracts.
In our examples, \prog{req} marks the clock cycle when \prog{address} sent by \prog{Top_Unsafe} or \prog{Top_Safe} is acknowledged on the channel.
The event \prog{res} marks the clock cycle when \prog{data} sent by the memory subsystem is acknowledged.
}

For memory without a cache, the expected behaviour is specified in a timing contract, encapsulated in the memory channel definition.
This contract requires that \prog{address} remain unchanging
and available for two clock cycles after \prog{req} is sent.
It also specifies that \prog{data} sent by memory must be available for one clock cycle after \prog{res} is received.

The timing contract is not satisfied by \prog{Top_Unsafe}, and \codename{} detects this at compile time.
In the HDL code for \prog{Top_Unsafe}, \prog{address} is sent during $[e_0, e_0+1)$, but the timing of acknowledgement is uncertain.
The output value is used during $[e_0+1, e_0+2)$, but when \prog{res} will
be received is unknown, as it depends on when the memory system responds.
As a result, it is unclear whether the next address was sent before the previous output was received and acknowledged.
Furthermore, the input address is modified during $[e_0+1, e_0+2)$, violating the requirement that the address remain unchanging for two cycles after acknowledgement.

The contract for memory with a cache is specified in the cache channel definition.
It requires that the \prog{address} sent by \prog{Top_Safe} remain available from the \prog{req} event until the next occurrence of \prog{res}, written as the lifetime \prog{(req, req->res)}.
Similarly, the \prog{data} sent by the memory subsystem has the lifetime \prog{(res, res->res+1)}.
As shown in Figure~\ref{fig:anvilChecker} (right), \prog{Top_Safe} satisfies this contract and is therefore deemed safe.

\nparagraph{Summary}{%
\codename{} is a general-purpose HDL that eliminates timing hazards.
It allows designers to declare timing contracts directly in the interface and provides higher-level abstractions to enforce those contracts.
The type system ensures that these contracts are respected.
}
\codename{} achieves this without sacrificing expressiveness. Dynamic contract definitions make it possible to design circuits with varying timing characteristics. It can capture timing characteristics precisely without introducing performance trade-offs such as additional latency.

\section{\codename{} HDL}

\label{sec:lang}

In this section, we give a tour of novel language primitives in \codename{} that are
relevant to timing safety.

\subsection{Channel}\label{subsec:chan}
\codename{} components communicate by message passing
through bidirectional \emph{channels},
which are akin to unbuffered channels in Go~\cite{go-channels}, where a
send and its corresponding
receive operations take place simultaneously.
Each \emph{channel type definition} in \codename{} describes a template for channels,
for example:
\begin{codebox}
\footnotesize
\begin{lstlisting}
chan mem_ch {
  left rd_req : (logic[8]@#1) @#2-@dyn,
  left wr_req : (addr_data_pair@#1),
  right rd_res : (logic[8]@rd_req) @#rd_req+1-@#rd_req+1,
  right wr_res : (logic[1]@#1) @#wr_req+1-@#wr_req+1
}
\end{lstlisting}
\end{codebox}

\nparagraph{Messages}{%
The definition specifies the different types of \emph{messages}
that can be sent and received over a channel {with two \emph{endpoints},
referred to as left and right, respectively}.
Each type of message is identified by a unique message identifier and annotated with its direction, which is \texttt{left} {(travelling left, i.e., from the right endpoint to the left endpoint)}
or \texttt{right} {(travelling right)}.%
}

\nparagraph{Message Contracts}{%
Each message is also associated with a
\emph{message contract}.
This contract specifies the data type of the message
and indicates the event after which the message content
is no longer guaranteed to remain unchanging and should,
therefore, be considered \textit{expired}.
Depending on the specified event of expiry,
a message contract can be
either static or dynamic.
For example, message \prog{rd_req} in the channel definition earlier
has a static contract:
It carries 8 bits of data, which expires 1 cycle after
the synchronization on the message takes place.
In contrast, message \prog{rd_res} has a dynamic contract:
It carries 8 bits of data which expires the next time
message \prog{rd_req} is sent or received.
}

\nparagraph{Sync Mode}{%
Each message has a \emph{synchronization mode} (\emph{sync mode} for short) for each side
of the communication.
The sync mode specifies the timing pattern for sending or receiving the message.
{%
In a message contract, the sync modes of both endpoints are specified in the format:
}
}

\begin{codebox}
\footnotesize
\begin{lstlisting}[language={}]
<left-endpoint-sync-mode>-<right-endpoint-sync-mode>
\end{lstlisting}
\end{codebox}

The default sync mode, \prog{@dyn}, specifies that a one-bit signal is used for run-time synchronization.
For example, in the channel definition, the message \prog{wr_req} uses this dynamic sync mode on both endpoints.
When static knowledge is available about when sending or receiving can occur, the sync mode can encode that information.
The left side of the message \prog{rd_req} has the \emph{static sync mode} \prog{@\#2}. This specifies that it must be ready to receive the message within at most two cycles after
the last time the message was received.
For the left endpoint, \codename{} statically checks that this constraint holds.
For the right endpoint, \codename{} uses this knowledge to check that whenever
\prog{rd_req} is sent, the receiver will be ready.
A sync mode can also be \emph{dependent}.
For example, both sides of \prog{wr_res} use \prog{@\#wr_req + 1}, meaning the message is sent and received exactly one cycle after \prog{wr_req}.

\subsection{Process}

Each \codename{} component is represented as a \emph{process},
defined with the keyword \texttt{proc}.
A process signature specifies a list
of endpoints to be supplied externally when the process is spawned.
The process body includes register definitions, channel instantiations,
other process instantiations, and threads.
\begin{codebox}
\footnotesize
\begin{lstlisting}
proc memory(ep1: left mem_ch, ep2: left mem_ch) {/* ... */}
\end{lstlisting}
\end{codebox}

\subsection{Thread}\label{sub:threads}
Each process contains one or more threads that execute concurrently.
Two types of threads are available: \emph{loops} and \emph{recursives}.

\nparagraph{Loops}{%
A loop is defined with \prog{loop \{ t \}}, where \texttt{t} is an \codename{} term (see
Section~\ref{subsec:term}).
This term can represent the parallel and sequential composition of multiple expressions.
Each time \texttt{loop\_term} completes execution,
the loop recurses back to the same behaviour.
For example, the code below increments
a counter every two cycles.
}

\begin{codebox}
\footnotesize
\begin{lstlisting}
loop { set counter := *counter + 1 >> cycle 1 }
\end{lstlisting}
\end{codebox}

\nparagraph{Recursives}{%
A recursive, defined with \prog{recursive \{ t \}}
generalizes loops to allow recursion before \texttt{t} completes.
Instead, recusion is controlled with \prog{recurse}.
As \texttt{t} can restart before it completes, multiple threads
may execute in an interleaving manner.
Such constructs are therefore
particularly useful for expressing simple \emph{pipelined} behaviours.
For example, the code below
pipelines the logic for handling the request.
{%
Specifically, it first waits to receive a request
message, then
performs two things in parallel:
1) handling the request, and
2) recursing (repeating the process from the beginning,
where it starts waiting for the next request) in the next
cycle.%
}
The direct pipelining support enabled by recursives
is comparable to
the pipelining support in Filament~\cite{nigamModularHardwareDesign2023}.

}

\begin{codebox}
\footnotesize
\begin{lstlisting}
recursive {
  let r = recv ep.rd_req >>
  { /* handle request */ };
  { cycle 1 >> recurse }
}
\end{lstlisting}
\end{codebox}

\subsection{Term}
\label{subsec:term}

Terms are the building block for describing computation and timing control of threads
in \codename{}.
Each term evaluates to a value (potentially empty) and the evaluation process
potentially takes multiple cycles.
In addition to literals and basic operators for computing (e.g., addition, xor, etc), notable
categories of terms include the following.

\nparagraph{Message Sending/Receiving}{%
The terms \prog{send e.m (t)} and \prog{recv e.m} send or receive
a specified message.
The evaluation completes when the message is sent or received.
}

\nparagraph{Cycle Delay}{%
The term \prog{cycle N} evaluates to an empty value after
    \prog{N} cycles and is used entirely for timing control.
}

\nparagraph{Timing Control Operators}{%
The \prog{>>} and \prog{;} operators are used for controlling timing. See Section~\ref{sub:wait}.
}

\subsection{Wait Operator}
\label{sub:wait}

The wait operator is a novel construct that enables
sequential execution by advancing to a time point.
In \prog{t1 >> t2}, the evaluation of the first term \prog{t1}
must be completed before the evaluation of the second term begins.
In contrast, \prog{t1; t2} initiates both term evaluations in parallel.
For example, \prog{set r := t} and \prog{set r := t; cycle 1} are equivalent,
since register assignment takes one cycle to complete.

This design not only provides a way to advance time by
explicitly specified numbers of cycles (e.g., \prog{cycle 2 >> ...}).
It also serves as an abstraction for managing and composing concurrent computations,
in a way similar to the async-await paradigm for asynchronous programming.
A term may represent computation that has not completed. Multiple
terms can be evaluated in parallel. When the evaluation result of a
term is needed, one can use \prog{>>} to wait for it to complete.
For example, the code below waits for messages from endpoints
\prog{ep1} and \prog{ep2} and processes the data concurrently.

\begin{codebox}
\footnotesize
\begin{lstlisting}
loop {
  let v1 = { let r = recv ep1.rd_req >> /* process r */ };
  let v2 = { let r = recv ep2.rd_req >> /* process r */ };
  v1 >> v2 >> ... /* now v1 and v2 are available */
}
\end{lstlisting}
\end{codebox}

\subsection{Revisiting the Running Example}
Figure~\ref{fig:anvilChecker} includes snippets
of \codename{} code for the running example
introduced in Section~\ref{sec:bg}.
The code demonstrates how \codename{} exposes
cycle-level control and supports expressing
dynamic timing behaviours.
The code uses the wait operator to control when and how
time is advanced.
It is clear from the source code when each operation takes place
relative to others.
In the bottom right timing-safe \codename{} code snippet,
for example, incrementing \prog{address} and
updating \prog{enq_data} take place at the same time (connected with
\prog{;}),
and sending of \prog{fifo.enq_req} starts
one cycle afterwards, when both register updates
complete.
Such timing control does not have to rely on fixed
number of cycles.
For example, the two register updates discussed above
take place after \prog{cache.res} is received,
which in turn takes place after \prog{cache.req}.
The exact numbers of cycles those operations vary during
run-time depending on the interaction between \prog{Top}
and \prog{Cache}.

Despite those dynamic timing behaviours that \codename{}
code can express, \codename{} is able to reason
about them and ensuring timing safety statically,
as we will discuss in detail next.

\section{Safety of \codename{} Programs}

\codename{}'s type system
ensures that each process
adheres to the contracts defined by the channels it uses.
The guarantee the type system provides is as follows: any well-typed process
in \codename{} can be composed with other well-typed processes without
\issuename{}s at run-time.
To provide such guarantees,
the type system associates each term with an abstract notion of a \emph{lifetime},
which, intuitively,
captures the time window in which its value is unchanging and meaningful.
Each register, likewise, is associated with a \emph{loan time}, which
describes when it is \emph{loaned}, i.e., needs to remain unchanged.
The abstractions of lifetime and loan time form the foundation for
ensuring safety in \codename{}.
Based on them, the type system checks for the following properties for a process ---
\begin{enumerate}
    \item \textbf{Valid Value Use:} Every use of a value falls in its associated lifetime.
    \item \textbf{Valid Register Mutation:} A register mutation does not take place
during its loan time.
\item \textbf{Valid Message Send:} The time window the data sent needs to be live for
(based on the timing contract) is covered by its associated lifetime. Additionally,
such time windows do not overlap for two send operations of the same message type.
\end{enumerate}

A formal presentation of the type system and the safety guarantees of \codename{} is
available in Section~\ref{sec:formal}.
We explain the intuition behind them in this section.

\subsection{Events and Event Patterns}
\codename{} reasons about events which correspond to
the times specific terms complete evaluation.
Note that such interesting events as sending and receiving of messages and
elapse of a number of cycles are naturally included, as the those operations are
all represented as terms (Section~\ref{subsec:term}).
\emph{Event patterns} can then be defined based on such events.
A basic event pattern is of the form $\timepattern{e}{p}$, which
consists of an existing event $e$ and a \emph{duration} $p$
and specifies the time when a condition specified in duration
$p$ is first satisfied after $e$.
The duration can be either static or dynamic.
A fixed duration specifies a fixed number of clock cycles, in the form of $\#N$.
A dynamic duration specifies a certain operation $\omega$, in which case
$\timepattern{e}{p}$ refers to when $\omega$ is first performed after $e$.
During run-time, a dynamic duration can correspond to variable numbers of cycles.
The typical example of a dynamic duration is the sending or receiving of
a specified message type through a channel.
In our discussion, this is represented as $\pi.m$, where $\pi$ is the
endpoint name and $m$ is the message identifier.
Multiple event patterns can be combined as a set of event patterns
$\{\timepattern{e_i}{p_i}\}_i$ to form a new event pattern,
which refers to the earliest event specified with each $\timepattern{e_i}{p_i}$.

\subsection{Lifetime and Loan Time}\label{LTandLT}

\nparagraph{Lifetime}{%
The lifetime represents the interval during which a value is expected to remain unchanging (constant).
\codename{} infers a lifetime for each value, represented by an interval $[e_{\sf start}, S_{\sf end})$, where
an event $e_{\sf start}$ and an event pattern $S_{\sf end}$ mark the beginning and end of the interval.
During run-time, the events $e_{\sf start}$ and $S_{\sf end}$ will correspond to specific clock cycles.
Since each signal carries a value, it inherently has an associated lifetime.
At any given instant, a signal is termed \emph{live} if it falls within its defined lifetime. Conversely, it is deemed \emph{dead}.
}

\nparagraph{Loan Time}{}Since signals and messages may source values from registers,
\codename{} tracks the intervals during which a register is loaned to a signal by associating each register with a loan time. The loan time of a register $r$ is a collection of intervals.
For each interval included in the loan time, $r$ should not be mutated.
\codename{} infers the lifetime for all associated values and the loan time for all registers.
Consider the example in Figure~\ref{fig:fooencrypt} (left) of a component named \texttt{Encrypt}.
This component performs encryption on the plaintext received through
the endpoint \prog{ch1} using
random noise obtained via the endpoint \prog{ch2}.
The following are examples of the lifetimes and loan times that \codename{} infers:

\begin{figure*}[]
\begin{minipage}{0.65\textwidth}
\begin{codebox}
\tiny
\begin{lstlisting}
chan encrypt_ch {
    left enc_req : (logic[8]@enc_res), right enc_res : (logic[8]@enc_req)
}
chan rng_ch {
    left rng_req : (logic[8]@#1), right rng_res : (logic[8]@#2)
}
proc Encrypt(ch1 : left encrypt_ch, ch2 : left rng_ch) {
    /* ... register definitions ... */
    loop {
     /*!\eventannotate{e_0}!*/ let /*!\typeannotatelocal{ptext}{e_1}{\timepattern{e_1}{\texttt{ch1.enc\_res}}}!*/ = recv ch1.enc_req;
     /*!\eventannotate{e_0}!*/ let /*!\typeannotatelocal{noise}{e_2}{\timepattern{e_2}{\#1}}!*/ = recv ch2.rng_req;
     /*!\eventannotate{e_0}!*/ let /*!\typeannotatelocal{r1\_key}{e_0}{\infty}!*/ = 25;
     /*!\eventannotate{e_0}!*/ /*!\typeannotatelocal{ptext}{e_1}{\timepattern{e_1}{\texttt{ch1.enc\_res}}}!*/ >>
     /*!\eventannotate{e_1}!*/ if ptext != 0 {
     /*!\eventannotate{e_1}!*/   /*!\typeannotatelocal{noise}{e_3}{\timepattern{e_2}{\#1}}!*/ >>
     /*!\eventannotate{e_3}!*/   set rd1_ctext := /*!\typeannotatelocal{(ptext \^\ r1\_key) + noise}{e_3}{\{\timepattern{e_2}{\#1}, \timepattern{e_1}{\texttt{ch1.enc\_res}}\}}!*/
     /*!\eventannotate{e_1}!*/ } else { rd1_ctext := /*!\typeannotatelocal{ptext}{e_1}{\timepattern{e_1}{\texttt{ch1.enc\_res}}}!*/ };
     /*!\eventannotate{e_1}!*/ cycle 1 >>
     /*!\eventannotate{e_5}!*/ set r2_key := /*!\typeannotatelocal{r1\_key \^\ noise}{e_6}{\timepattern{e_2}{\#1}}!*/;
     /*!\eventannotate{e_5}!*/ let ctext_out = /*!\typeannotatelocal{*rd1\_ctext \^\ *r2\_key}{e_5}{\timepattern{e_9}{\texttt{ch1.enc\_req}}}!*/;
     /*!\eventannotate{e_5}!*/ send ch2.rng_res(/*!\typeannotatelocal{*r2\_key}{e_5}{\timepattern{e_8}{\#2}}!*/) >>
     /*!\eventannotate{e_8}!*/ send ch1.enc_res(/*!\typeannotatelocal{ctext\_out}{e_8}{\timepattern{e_9}{\texttt{ch1.enc\_req}}}!*/) >>
     /*!\eventannotate{e_9}!*/ send ch1.enc_res(/*!\typeannotatelocal{r1\_key}{e_9}{\infty}!*/)
    }
}
\end{lstlisting}
\end{codebox}
\end{minipage}
\hfill
\begin{minipage}{0.33\textwidth}
\centering
\includegraphics[width=.7\linewidth]{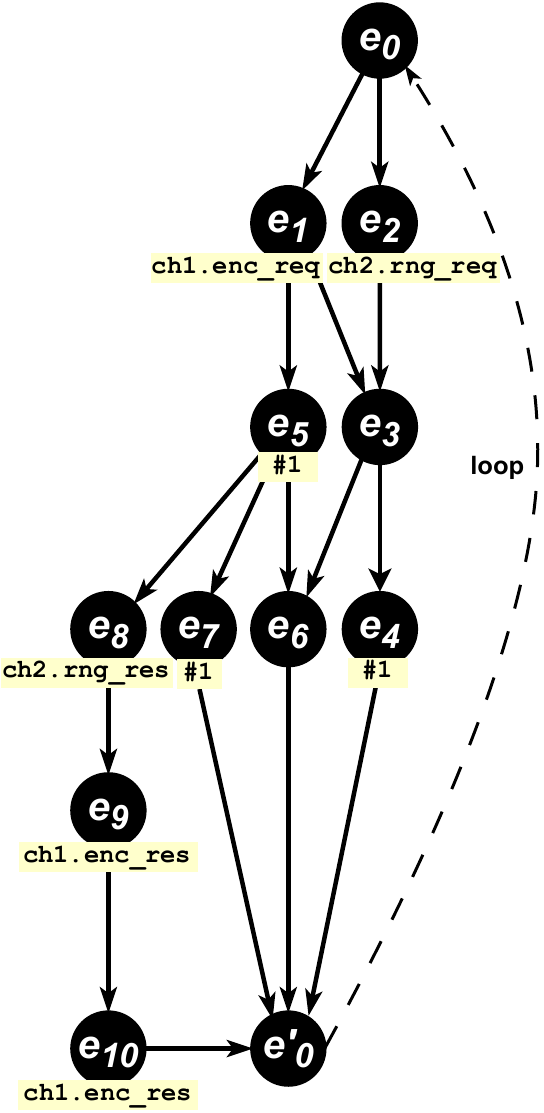}
\end{minipage}
\caption{Left: \texttt{Encrypt} in \codename{}, annotated with timing
information. Each blue-shaded annotation marks the event corresponding
to the time a term evaluation starts.
Each yellow-shaded annotation marks the inferred lifetime associated with
the red-circled term next to it.
Right: Event graph corresponding to \texttt{Encrypt}.
Branch-related constructs which exist in
the event graph actually used in the type system
are omitted for brevity.
The operations associated with
some of the events are presented in yellow labels.}
\label{fig:fooencrypt}
\end{figure*}

\begin{itemize}

\item The signal \prog{ptext} is bound to a
message identified by \prog{enc_req}
received on the endpoint \prog{ch1}. Its lifetime is inferred from the channel type definition as $[e_1, \timepattern{e_1}{\texttt{ch1.enc\_res}})$,
where $e_1$ is the event of the message being received.

\item The signal \prog{r1\_key} is a constant literal and therefore has an \emph{eternal} lifetime,
represented with $\infty$ as its end event.
i.e., it can always be used.

\item The signal \prog{ctext\_out} is used as a value sent as a message from the endpoint \prog{ch1}.
Its inferred lifetime begins at the evaluation of the term, represented as $e_5$,
and extends until the message on \prog{ch1} expires, which is
$\timepattern{e_9}{\texttt{ch1.enc\_req}}$, where
$e_9$ is the event corresponding to the completion of the message sending.
Therefore, the lifetime is $[e_5, \timepattern{e_9}\texttt{ch1.enc\_req})$.

\item The signal \prog{(ptext \^\ r1\_key) + noise} has a lifetime that is the
intersection of the lifetimes of \prog{ptext}, \prog{r1\_key}, and \prog{noise},
$[e_3, \{\timepattern{e_2}{\#1}, \timepattern{e_1}{\texttt{ch1.enc\_res}}\})$.

\item The register \prog{rd2\_key} is loaned by a message sent through the endpoint \prog{ch2} and the signal \prog{ctext\_out}.
Based on the specified timing in the channel type definition \prog{rng\_ch},
the lifetime of the message is $[e_5, \timepattern{e_8}{\#2})$, where
$e_8$ is the event of the message sending completion.
Therefore, \texttt{rd2\_key} has an inferred loan time $[e_5, \timepattern{e_9}\texttt{ch1.enc\_req}) \cup
[e_5, \timepattern{e_8}{\#2})$.
\end{itemize}

See Figure~\ref{fig:fooencrypt} (left) for more examples of inferred lifetimes.

\subsection{Event Graph}
\label{subsec:event-graph}

Events are related to one another by their associated operations.
For example, an event $e_a$ may be precisely two cycles after another event $e_b$.
As another example, $e_a$ can refer to the completion of a \emph{specific} message that
\emph{starts} at $e_b$.
In general, events and their interrelationships
form a directed acyclic graph (DAG), with each
node being an event labelled with its
associated operation.
We call such a DAG an \emph{event graph}.
\texttt{Encrypt} in Figure~\ref{fig:fooencrypt} (left), for example, has an event graph as shown
in Figure~\ref{fig:fooencrypt} (right).
The event graph captures the events in one loop iteration only, with
event $e_0$ representing the start of a loop iteration.
The event $e^\prime_0$ corresponds to $e_0$ of the next loop iteration.

An event graph encodes sufficient information to capture all possible timing
behaviours in run-time.
Intuitively, once we replace each non-cycle operation label (e.g., those associated with
$e_1, e_2, e_8, e_9$, and $e_{10}$ in Figure~\ref{fig:fooencrypt} (right)) with
a cycle number that represents the actual amount of time taken to complete the message passing,
we can deterministically obtain the exact time (in cycles) each event occurs.

\subsection{Safety Checks}
\label{safetychecks}

\nparagraph{Building Blocks: $\leq_G$ and $\subseteq_G$}{%
Based on an event graph $G$,
\codename{} compare pairs of events as to the order in which they occur during run-time.
In particular, \codename{} decides
if an event $e_a$ always occurs no later than
another event $e_b$, denoted as $\nolater{G}{e_a}{e_b}$.
The simple scenario is when a path exists from $e_a$ to $e_b$ in $G$
and we directly have $\nolater{G}{e_a}{e_b}$.
More complex scenarios
involve events with no paths between them, which \codename{}
handles by considering the ``worst'' cases time gap between when
the two events are reached.
For example,
we have $\nolater{G}{e_5}{e_4}$, as even in the worst case (receiving \prog{ch2.rng\_req} takes
0 cycles), $e_4$ and $e_5$ still occur at the same time.
We naturally extend the definition of $\leq_G$ to cover
event patterns and reuse the notation $S_a \leq_G S_b$.

With $\leq_G$, the \codename{} type system can decide that
an interval $[e_a, S_a)$ is always fully within another interval $[e_b, S_b)$,
denoted $[e_a, S_a) \subseteq_G [e_b, S_b)$,
if $\nolater{G}{e_b}{e_a}$ and $\nolater{G}{S_a}{S_b}$.
It then decides if the lifetimes and the loan times comply with
the three types of constraints.
We use the example in Figure~\ref{fig:fooencrypt} to explain them below.
}

\nparagraph{Valid Value Use}{%
\codename{}'s type system verifies that events
at which a signal is used are within its defined lifetime.
A use of \prog{ptext} occurs at $e_1$ in
the expression \prog{if ptext != 0 \{ ... \}}, where it has a
a lifetime of $[e_1, \timepattern{e_1}{\texttt{ch1.enc\_res}})$.
It requires \prog{ptext} to be live for one cycle, i.e., in $[e_1, \timepattern{e_1}{\#1})$.
\codename{} checks that $[e_1, \timepattern{e_1}{\#1}) \subseteq_G [e_1, \timepattern{e_1}{\texttt{ch1.enc\_res}})$,
which holds in this case.
Hence \codename{} determines that \prog{ptext} is guaranteed to be live during this read.

However, in \prog{rd1\_ctext := (ptext \^\ r1\_key) + noise},
the signal \prog{(ptext \^\ r1\_key) + noise} cannot be statically guaranteed to be live.
In this case, \codename{} compares its lifetime, $[e_3, \{\timepattern{e_2}{\#1},
\timepattern{e_1}{\texttt{ch1.enc\_res}}\})$ with the time when it is used,
$[e_3, \timepattern{e_3}{\#1})$ {(the assignment starts at event $e_3$ and takes
one cycle to complete).}
It cannot obtain $\nolater{G}{\timepattern{e_3}{\#1}}{\{\timepattern{e_2}{\#1},\timepattern{e_1}{\texttt{ch1.enc\_res}}\}}$.
Intuitively, if it takes more cycles to receive \prog{ch1.enc\_req} ($e_1$) than \prog{ch2.rng\_req} ($e_2$),
\prog{noise} will already be dead at $e_3$ when the assignment happens.

}

\nparagraph{Valid Register Mutation}{%
\codename{} ensures that
each register value remains constant during its loan time.
For the example in Figure~\ref{fig:fooencrypt}, the loan time for \prog{r2\_key} is
$[e_5, \timepattern{e_9}\texttt{ch1.enc\_req}) \cup
[e_5, \timepattern{e_8}{\#2})$.
To determine if
\prog{r2\_key} is still loaned when the assignment \prog{r2_key := r1\_key \^\ noise}
takes place,
\codename{} checks if $[e_5, \timepattern{e_7}{\#1})$ is guaranteed not to be fully
covered by any interval in its loan time, {i.e., for every $[e^\prime, S^\prime)$ in
the loan time, either $S <_G S^\prime$ or $e^\prime <_G e$ must hold.}
{
Here, $e_7$ is the event that corresponds to the completion of the assignment, which is
exactly one cycle after $e_5$, which corresponds to when the assignment starts.}
In other words, $e_5$ and $e_7$ are adjacent cycles in which the register can carry different values.
If an interval in the loan time may contain both $e_5$ and $e_7$, at run-time during the interval the
register value may change.
In the example, $[e_5, \timepattern{e_8}{\#2})$ potentially (surely in this case)
fully covers $[e_5, \timepattern{e_7}{\#1})$, hence this assignment conflicts with the loan time
of \prog{r2\_key} and is not allowed by \codename{}.
Intuitively, a value sourced from \prog{r2\_key} is sent through \prog{ch2.rng\_res} at $e_5$,
which requires it to be live until two cycle after the send completes.
However, the value \prog{r2\_key} already changes one cycle after $e_5$.

}

\nparagraph{Valid Message Send}{%
In the example in Figure~\ref{fig:fooencrypt}, the term \prog{send ch1.enc\_res(r1\_key)} attempts
to send a new message before the previous \prog{enc\_res} message sent by the endpoint \prog{ch1} has expired.
During run-time on the other end of channel,
this can lead to signals received through \prog{enc\_res} to change,
violating the message contract.
\codename{} detects such violations by examining
whether the required lifetimes of the two send operations are disjoint.
The example violates such constraints as
$[e_8, \timepattern{e_9}{\texttt{ch1.enc\_req}})$ and
$[e_9, \timepattern{e_{10}}{\texttt{ch1.enc\_req}})$ are overlapping.

\codename{} also checks that the lifetimes of sent signals
cover the required lifetime specified by the message contract.
For example, the send through \prog{ch1.enc\_res} at $e_9$ checks that the lifetime
of \prog{r1\_key} covers the required lifetime $[e_9, \timepattern{e_{10}}{\texttt{ch1.enc\_req}})$.
In this case, this check passes as $[e_9, \timepattern{e_{10}}{\texttt{ch1.enc\_req}}) \subseteq_G
[e_9, \infty)$.
}

\subsection{Formalization}

\label{sec:formal}

\begin{figure}[t]
    \centering
    \footnotesize
    \begin{align*}
    & \text{process definition} \quad & P &::= \texttt{proc}\ p(\pi, \ldots) \{ B \} \\
    &\text{process body} \quad & B &::= \emptyset \mid \texttt{reg}\ r : \delta; B \mid \texttt{ch}\ c(\pi, \pi); B \\
    & & & \quad \mid \texttt{spawn}\ p(\pi, \ldots); B \mid \texttt{loop}\ \{t\}\ B \\
    & \text{term} \quad & t &::= \texttt{true} \mid \texttt{false} \mid \texttt{()} \mid
    \texttt{cycle } N \mid x \mid *r \\
    & & & \quad \mid t \Rightarrow t \mid \texttt{let}\ x = t\ \texttt{in}\ t \mid \texttt{ready} \ (\pi.m) \\
    & & & \quad \mid \texttt{if}\ x\ \texttt{then}\ t\ \texttt{else}\ t \mid \texttt{send}\ \pi.m(x) \\
    & & & \quad \mid \texttt{recv}\ \pi.m \mid r := t \mid t \star t \mid \pentagon t 
    \end{align*}
    \begin{multline*}
    \delta \in \text{data-types}  \quad
    \star \in \text{binary-operators} \quad
    \pentagon \in \text{unary-operators} \\ 
    \pi \in \text{endpoints} \quad
    x \in \text{identifiers} \quad
    r \in \text{registers} \quad
    m \in \text{messages} \\
    c \in \text{channels}\quad p \in \text{processes} \quad
    N \in \mathbb{N}
    \end{multline*}
\caption{\codename{} syntax.}
\label{fig:syntax}
\end{figure}

Figure~\ref{fig:syntax} presents the syntax of \codename{}.
\codename{}'s type system guarantees that any well-typed \codename{} program is timing-safe.
Due to space limits, we leave the formal details of the semantics, the type system of \codename{}, the safety definitions, and proofs
to Appendices~\ref{appx:details} and \ref{appx:proofs}.

\section{Implementation}
\label{sec:impl}

We have implemented \codename{} in OCaml. The \codename{} compiler
performs type checking on \codename{} code and generates synthesizable
SystemVerilog.
We have publicly released the compiler at
\url{https://github.com/jasonyu1996/anvil}.

The compiler uses
the event graph
as an intermediate representation (IR)
throughout the compilation process.
It constructs an event graph
from the concrete syntax tree of the \codename{} source code,
performs type checking on it, and
lowers it to SystemVerilog.
Optimizations are applied to the event graph both
before and after type checking.
Since event graph construction and type checking follow the type system
in a straightforward manner, we focus on the optimization and lowering strategies
in this section.

\subsection{Event Graph Optimizations}
\label{subsec:opt}
\begin{figure}[t]
    \centering
    \includegraphics[width=\linewidth]{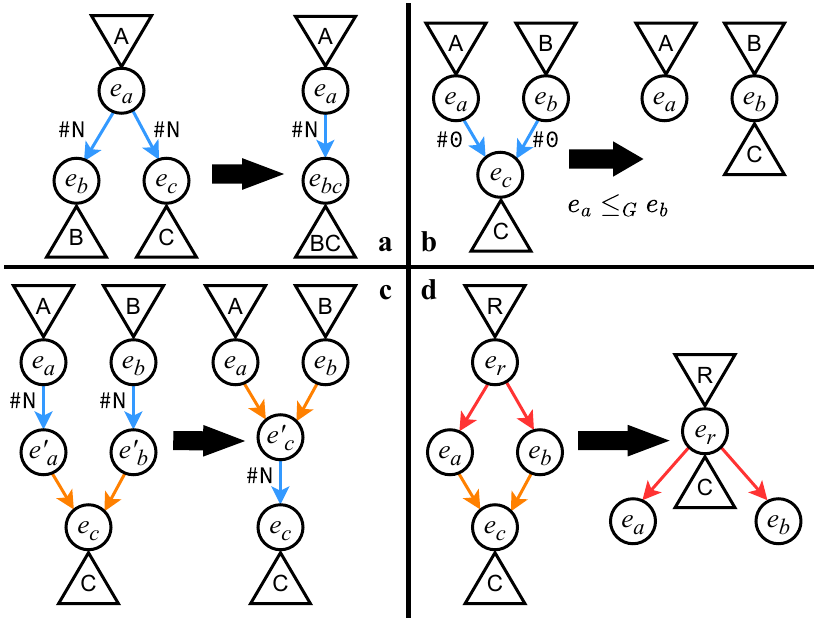}
    \caption{Examples of event graph optimizations. {The event graphs
    are simplified for illustration purposes. Blue edges represent fixed delays.
    }}
    \label{fig:optimizations}
\end{figure}

Optimizations aim to reduce the number of events in the event graph while
keeping its semantics unchanged.
The \codename{} compiler performs optimizations in \emph{passes}, with each pass
applying a specific optimization strategy.
Figure~\ref{fig:optimizations} shows examples of such optimization passes.
{
The figure illustrates simplified event graphs during optimization.
Edge labels (including
their colours illustrated in the figure) describe the timing relationships between
events (nodes in the figure).
A blue edge from $e_a$ to $e_b$ represents that
$e_b$ waits for a fixed delay after $e_a$, with the number of cycles indicated in
\texttt{\#N}).
When an event waits on multiple other events, i.e., with multiple inbound
blue edges, it occurs at the latest
of the specified time points.
Red edges represent branching. When $e_r$ has red edges to both $e_a$
and $e_b$, either of them, but not both, occurs in the same cycle as $e_r$.
Orange edges in turn join branches: When $e_a$ and $e_b$ both have
orange edges to $e_c$, $e_c$ occurs in the same cycle when either of them occurs.
Triangles represent sets of edges.
Recall that events represent abstract time points.
In a concrete run, they occur in specific cycles (or are never reached).}
In general, two events can be merged if they always occur
at the same time.
Many of the optimization passes are based on identifying such cases.

\nparagraph{(a) Merging identical outbound edge labels}{%
   This optimization pass merges outbound edges of an event $e_a$ that
   share the same label.
   For example, edges labelled with \texttt{\#N} going to $e_b$ and $e_c$.
   The events those edges connect to are merged.
   A shared label implies that they occur at identical delays from the parent event.
}

\nparagraph{(b) Removing unbalanced joins}{%
    This optimization pass removes an $e_c$ with two
    predecessors when either of its predecessors
    ($e_b$) always occurs no earlier than the other ($e_a$), i.e.,
    $e_a \leq_G e_b$.
    In this case, $e_c$ is unnecessary, and its outbound edges
    are migrated to $e_b$.
}

\nparagraph{(c) Shifting branch joins}{
A \emph{branch join} is an event $e_c$ that joins two branches.
When the ending events of the two branches $e^\prime_a$ and $e^\prime_b$
are both derived with
$N$ cycles delay after their predecessors $e_a, e_b$,
and have no associated actions (e.g., register assignments or message sends/receives),
the event $e_c$ that joins the two can be shifted earlier.
Instead of delaying by $N$ cycles and then joining,
the branches can join first into $e^\prime_c$ and then delay by $N$ cycles.
}

\nparagraph{(d) Removing branch joins}{%
If an event $e_c$ joins two branches where the ending events $e_a$ and $e_b$
are also the first events of their branches,
and both share the same predecessor $e_r$,
then $e_c$ can be merged into $e_r$.
This means that if two branches take no delay, their joining event can be merged into the predecessor.
All actions of the joining event are then performed in the predecessor event.
}

\subsection{Code Generation}

The \codename{} compiler maps each \codename{} process to
a SystemVerilog module. For each process, it generates
module input/output ports for channel communication
and a finite state machine (FSM) for control flow
based on the event graph. Note that the compiler
generates \emph{no} extra code for maintaining lifetimes or
enforcing timing safety as it reasons about lifetimes statically
and guarantees timing safety through static type checking.
As such, they incur no overhead in the generated hardware design.

\nparagraph{Message Lowering}{%
Each message in an endpoint maps to up to three module ports:
\texttt{data}, \texttt{valid}, and \texttt{ack}.
The \texttt{data} port carries the communicated data, while
\texttt{valid} and \texttt{ack} are handshake ports used
in the synchronization.
The compiler only generates both \texttt{valid} and \texttt{ack}
when the specified synchronization mode is dynamic for both
the sender and the receiver.
If the synchronization mode for either side is static or dependent,
the compiler omits the corresponding port (\texttt{valid} for the sender
and \texttt{ack} for the receiver).
In particular, both handshake ports are omitted for a synchronization mode that
is not dynamic on either side, leaving \texttt{data} as the only port generated.
}

\nparagraph{FSM Generation}{%
The compiler generates the FSM based on the event graph structure.
For each event, it uses a one-bit wire \texttt{current} to indicate if the
event has been reached.
For some events, the compiler also generates registers to record the current state.
Such events include:
(a) Joins: which predecessors have been reached;
(b) Cycle delays: cycle count;
(c) Send/receive events (only those with dynamic synchronization modes):
    whether the message has been sent or received.
}

\section{Evaluation}
\label{sec:eval}

We aim to answer three questions through evaluation:

\begin{enumerate}
    \item \textbf{Expressiveness:} Can \codename{} express diverse hardware designs, without incurring any latency\footnote{Latency refers to clock cycle latency and not propagation delay.} overhead?
{\item \textbf{Safety:} Can \codename{} assist the designer to express and meet the implicit timing contracts?}
\item \textbf{Practicality:} What overheads do \codename{}-generated hardware designs incur in synthesis?

\end{enumerate}

\subsection{Expressiveness}

{SystemVerilog supports describing circuits with arbitrary latencies. To assess expressiveness, we evaluate designs created in \codename{}
against open-source hardware components written in SystemVerilog.
We also compare \codename{} with Filament~\cite{nigamModularHardwareDesign2023}, which provides specialized abstractions for static pipelines only.
}

\begin{table*}
\centering
\caption{Summary of area and power footprints of Anvil and baseline designs in SystemVerilog and Filament. \emph{SV} stands for SystemVerilog and
\emph{dyn} indicates dynamically varying cycle latencies.}
\label{tab:area-power}
\resizebox{\textwidth}{!}{
\begin{tabular}{|l|ll|ll|ll|ll|}
\hline
\multicolumn{1}{|c|}{\multirow{2}{*}{\textbf{Hardware Designs}}} & \multicolumn{2}{c|}{\textbf{Area ($\text{um}^2$)}}                                           & \multicolumn{2}{c|}{\textbf{Power ($\text{mW}$)}}                                          & \multicolumn{2}{c|}{\textbf{$f_\text{max}$ (MHz, $\pm 50$)}}                                           & \multicolumn{2}{c|}{\textbf{Latency (cycles)}}                                        \\ \cline{2-9}
\multicolumn{1}{|c|}{}                                           & \multicolumn{1}{c|}{\textbf{Baseline}} & \multicolumn{1}{c|}{\textbf{Anvil}} & \multicolumn{1}{c|}{\textbf{Baseline}} & \multicolumn{1}{c|}{\textbf{Anvil}} & \multicolumn{1}{c|}{\textbf{Baseline}} & \multicolumn{1}{c|}{\textbf{Anvil}} & \multicolumn{1}{c|}{\textbf{Baseline}} & \multicolumn{1}{c|}{\textbf{Overhead}} \\ \hline
\textbf{FIFO Buffer} (SV)                                              & \multicolumn{1}{l|}{690}               & 674 ($-2\%$)                                 & \multicolumn{1}{l|}{1.434}             & 1.403 ($-2\%$)                               & \multicolumn{1}{l|}{4062}              &4156                                & \multicolumn{1}{l|}{dyn}                 & 0                                   \\ \hline
\textbf{Spill Register} (SV)                                     & \multicolumn{1}{l|}{165}               & 171 ($3\%$)                           & \multicolumn{1}{l|}{0.459}             & 0.469 (2\%)                             & \multicolumn{1}{l|}{5187}              &5375                                & \multicolumn{1}{l|}{dyn}                 & 0                                   \\ \hline
\textbf{Passthrough Stream FIFO} (SV)                           & \multicolumn{1}{l|}{679}               & 679 ($0\%$)                               & \multicolumn{1}{l|}{1.239}             & 1.264 ($2\%$)                        & \multicolumn{1}{l|}{4093}              & 3625                                & \multicolumn{1}{l|}{1}                 & 0                                   \\ \hline
\textbf{CVA6 Translation Lookaside Buffer} (SV)                                                & \multicolumn{1}{l|}{5561}              & 5611 ($0\%$)                        & \multicolumn{1}{l|}{5.813}             & 5.835 ($0\%$)                     & \multicolumn{1}{l|}{2468}              &2406                                & \multicolumn{1}{l|}{dyn}                 &0                                   \\ \hline
\textbf{CVA6 Page Table Walker}  (SV)                                             & \multicolumn{1}{l|}{499}               & 561 ($12 \%$)                         & \multicolumn{1}{l|}{0.649}             & 0.676 ($4 \%$)                        & \multicolumn{1}{l|}{3531}              & 3281                                & \multicolumn{1}{l|}{dyn}               & 0                                 \\ \hline
\textbf{AES Cipher Core} (SV)                                         & \multicolumn{1}{l|}{9096}              & 9090 ($0\%$)                                & \multicolumn{1}{l|}{0.793}             & 0.972  ($22\%$)                        & \multicolumn{1}{l|}{781}              & 1229                                & \multicolumn{1}{l|}{dyn}             & 0                               \\ \hline
\textbf{AXI-Lite Demux Router}  (SV)                                  & \multicolumn{1}{l|}{1318}              & 1469 ($11\%$)                         & \multicolumn{1}{l|}{1.351}             & 1.385 ($2\%$)                         & \multicolumn{1}{l|}{2437}              & 2125                                & \multicolumn{1}{l|}{dyn}               & 0                                 \\ \hline
\textbf{AXI-Lite Mux Router}  (SV)                                    & \multicolumn{1}{l|}{1448}              & 1633 ($12 \%$)                        & \multicolumn{1}{l|}{1.336}             & 1.324 ($0\%$)                             & \multicolumn{1}{l|}{2406}              & 2187                                & \multicolumn{1}{l|}{dyn}               & 0                                \\ \hline
\multicolumn{9}{|c|}{\textbf{Average overhead compared with SystemVerilog baselines: $\text{Area} = 4.50\%, \text{Power} = 3.75\%$}} \\ \hline
\textbf{Pipelined ALU}  (Filament)                                         & \multicolumn{1}{l|}{501}                  & 404  ($-19\%$)                                   & \multicolumn{1}{l|}{0.658}                  &0.678 ($3 \%$)                                     & \multicolumn{1}{l|}{3312}                  &4675                                     & \multicolumn{1}{l|}{1}                  &0                                     \\ \hline
\textbf{Systolic Array}  (Filament)                                         & \multicolumn{1}{l|}{2522}                  &2434 ($-3 \%$)                                    & \multicolumn{1}{l|}{2.533}                  &2.808 ($10 \%$)                                    & \multicolumn{1}{l|}{2437}                  &2862                                     & \multicolumn{1}{l|}{1}                  &0                                     \\ \hline
\multicolumn{9}{|c|}{\textbf{Average overhead compared with Filament baselines: $\text{Area} = -11.0\%, \textbf{Power} = 6.5\%$}} \\ \hline
\end{tabular}
}
\end{table*}
{
\nparagraph{Common Cells Benchmarks}{%
\codename{} is designed to be a general-purpose HDL. To test this, we implemented various hardware components with different behaviours. Specifically, we implemented a first-in first-out (FIFO) buffer, a spill register,
and a passthrough stream FIFO (which allows read and write in the same cycle). These are taken from the Common Cells IP  and are highly optimised designs for synthesis~\cite{commoncellsgithub}.
With \codename{}, we replicated these designs while ensuring identical functional behaviour through unit tests. Importantly, \codename{} is able to express their dynamic behaviour \emph{without introducing any latency overhead}.}}

\nparagraph{CVA6 MMU}{We implemented the translation lookaside buffer (TLB) and page table walker (PTW), which together form the core of the memory management unit (MMU) in the CVA6 RISC-V core~\cite{zarubaCostApplicationclassProcessing2019} These units are highly sensitive to dynamic latencies, which cannot be captured by static contracts.
For example, the PTW performs a page table walk across up to three levels of pages, and can therefore respond to requests with varying latencies. \codename{} replicates the same functional behaviour (verified with baseline RISC-V smoke tests) \emph{without incurring any cycle-level latency overhead over the baselines}.}

{\nparagraph{OpenTitan AES Accelerator ~\cite{opentitanaes}}{We implemented the unmasked AES cipher core from OpenTitan. This core supports encryption, decryption, and on-the-fly key generation for AES-128 and AES-256. We verified its functional behaviour using unit tests for encryption and decryption of plaintext.
The core has a clock-cycle latency proportional to the number of encryption rounds, and it flushes its state during operation. These characteristics make the latency dynamic, and \codename{} is able to replicate this behaviour.
The original AES core uses an S-box implementation intended for LUT mapping. To stay consistent with this design choice, we used the baseline S-box IP optimized for LUT realization.}

\nparagraph{AXI-Lite Routers}{\codename{} abstracts communication interfaces using channels. To demonstrate the utility of this abstraction in real-world components, we implemented the AXI-Lite demux router and AXI-Lite mux router with fair arbitration.
The AXI protocol itself is designed to provide a channel-like interface between master and slave components. We verified the correctness of our implementations using unit tests with configurations of 8 slaves and 1 master, and vice versa.
These routers can be composed into an AXI crossbar according to the desired configuration. With \codename{}, we replicated the same functional behaviour while abstracting away the complexity of handling transaction requests from the user. As in all our experiments, this design also does not incur any additional latency overhead.}

\nparagraph{Pipelined Designs}{Lastly, to demonstrate the ability of \codename{} recursives (Section \ref{sub:threads}) to express static pipelined designs, we implemented a pipelined ALU and a pipelined systolic array.
We compared these implementations against hardware designs generated by Filament.
The evaluation shows that \codename{} allows for expressing such designs without incurring any additional penalty.}}

\tinyskip

\noindent\fbox{
\begin{minipage}{.95\linewidth}
   \textbf{Takeaway.} \codename{} provides
   cycle-level timing control and precise expression of dynamic latency, with no additional cycle latency and throughput overhead.
\end{minipage}
}

\subsection{Safety}
During our evaluation, we observed issues with the stream FIFO. According to the IP documentation, the design goal is clear:
Reads are allowed only when the FIFO is not empty, and writes only when it is not full. Additionally, if there is a read and write request in the same cycle and the FIFO is full, it should still allow the write.

However, we noticed that the original FIFO, even with a handshake interface, does not actually prevent writes at all. Instead, it relies on warning assertions (SVA) to alert designers if they run into such cases. This means that unless the design hits a specific overflow condition, no assertion is raised. Moreover, once the overflow happens, there is no further assertion until the FIFO again reaches its full depth.

This creates a gap between the documented contract and the actual behaviour. The design does contain possible timing hazards and effectively pushes the responsibility onto the designer to avoid them. In contrast, Anvil enforces these contracts directly, and as we observe, does so without incurring significant overhead.
There are several such examples of timing hazards in open-source IPs, where enforcement is either left to the designer or sometimes not handled at all. We discuss several such instances in Appendix \ref{appx:eval}.

\subsection{Practicality}

To evaluate the practicality of the generated designs, we synthesized all of them on a commercial 22 nm ASIC process. This shows how well \codename{} designs scale during synthesis compared to SystemVerilog, which is widely regarded as the most efficient option for practical hardware. We then provide a detailed analysis of the sources of overhead and efficiency in these designs. Table \ref{tab:area-power} summarizes the resource consumption of circuits generated with \codename{}.

{
\nparagraph{Setup}{We report the area, power, the maximum frequencies ($f_\text{max}$) at which designs
do not violate slack requirements,
and the clock cycle latency of the baseline.
The area and power are reported at $\min (f_\text{max}(\text{\codename{}}), f_\text{max}(\text{baseline}))/2$.}

\nparagraph{Propogation Delay}{The maximum frequency evaluation shows that \codename{} is able to synthesize circuits that are not worse than the baseline in supporting higher frequencies.
This is primarily because the critical path is the same in both designs. The additional propagation delay only comes from the extra combinational logic introduced by code generation. For pipelined designs, \codename{} achieves higher maximum frequency than the baseline.
}

\nparagraph{Area}{\codename{} provides constructs that implicitly generate state machines as efficiently as handwritten ones. This is reflected in the area overheads when compared against handwritten baselines. The overhead in non-combinational area across all designs is equivalent to, or in some cases even lower than, the baseline implementations.}

For example, in the case of the AXI router, the observed overhead for the AXI demux (1469 vs. 1318) arises entirely from the FIFO component. The FIFO is required to preserve the ordering of transactions on the AW/AR channels relative to their corresponding W/B/R channels. The router instantiates three FIFO queues in total. Each FIFO contributes roughly 45 units of area overhead, as the select signal width is only 3 bits. However, as the data width increases, the relative overhead becomes demagnified. This trend is evident in the 32-bit FIFO buffer results reported in Table~\ref{tab:area-power}.

A similar observation holds for the PTW. Here, the non-combinational area is comparable (330 vs. 352), while the combinational area shows a modest gap (168 vs. 208). This difference essentially reflects a fixed cost of \codename{}’s code generation. As a result, the relative overhead is more pronounced for small-area designs but  negligible for larger ones.

\nparagraph{Power}{The power overhead in \codename{} arises primarily from bundling signals and flattening data structures. In this representation, the synthesis toolchain may treat the entire bundle as active, even when only a portion is in use. Consequently, switching activity, and thus dynamic power, increases with datapath width, as observed in the AES cipher core.
At the same time, the \codename{} compiler can reduce leakage power because all signals are explicitly connected, leaving none floating. Additionally, register assignments for explicitly declared registers occur only when the corresponding event is triggered. This behaviour implicitly provides clock gating for some register writes.
}

\nparagraph{Summary}{%
Overall, Table~\ref{tab:area-power} shows that \codename{} achieves area efficiency on par with handwritten SystemVerilog, with overheads typically within $12\%$ and averaging $4.50\%$.
Power overheads are more noticeable in wide datapath designs (e.g., the AES cipher core) due to increased switching activity, but remain modest overall
(averaging $3.5\%$).
The maximum frequencies are generally on par with handwritten SystemVerilog, and in pipelined cases, even exceed the baseline.
Importantly, none of the designs introduce extra cycle latency.
}
}

\tinyskip

\noindent\fbox{
\begin{minipage}{.95\linewidth}
   \textbf{Takeaway.}
   \codename{} is practical for creating real-world hardware designs
   with minimal area/power overheads and seamlessly integrates into existing SystemVerilog designs.
\end{minipage}
}

\section{Related Work}

\label{sec:related-work}

\nparagraph{Timing-Oblivious HDLs}{%
The industry-standard HDLs, SystemVerilog~\cite{18002017IEEEStandard2018}
and VHDL~\cite{VHDLRef}, describe hardware behaviours with dataflows involving registers and wires
within single cycles.
This abstract model equips them with low-level expressiveness but is not conducive to
time-related reasoning, causing such problems as \issuename{}s.
Embedded HDLs~\cite{bachrachChiselConstructingHardware2012,spinalhdl,myhdl,rayHardcamlOCamlHardware2023}
use software programming languages
for hardware designs for their better parameterization and abstraction capabilities.
They follow the same single-cycle model as in SystemVerilog and VHDL.
Bluespec SystemVerilog~\cite{bsv,bourgeatEssenceBluespecCore2020} provides an abstraction
of hardware behaviours with sequential firing of atomic rules.
It is still limited to describing single-cycle behaviours and does not provide timing safety.
Higher-level HDLs, high-level synthesis (HLS) languages, and accelerator description languages (ADLs)
~\cite{zagieboyloPDLHighlevelHardware2022,TLVerilog,xls,spade,systemc}
specialize in specific applications and
abstracts away cycle-level timing and the distinction between stateless signals and registers.
}

\nparagraph{Timing-Aware HDLs}{%
Filament~\cite{nigamModularHardwareDesign2023}
achieves timing safety with timeline types which only support statically fixed delays.
As a result, it is limited to designs with static timing behaviours.
HIR~\cite{majumderHIRMLIRbasedIntermediate2023}
is an intermediate representation (IR) for describing accelerator designs.
It introduces time variables to specify timing, and allows specifying
a static delay for each function to indicate when it returns.
HIR abstracts away the distinction between
signals and registers and does not capture the notion of lifetimes and only supports static timing behaviours.
Piezo~\cite{kimUnifyingStaticDynamic2024} is an IR that supports
specifying both static and dynamic timing through timing guards.
}

\nparagraph{Hazard Prevention}{%
BaseJump~\cite{BaseJump} and Wire Sorts~\cite{wireSort} are type systems deigned
to identify combinational loops, a separate concern than \issuename{}s.
ShakeFlow~\cite{hanShakeFlowFunctionalHardware2023} proposes a dynamic control
interface to prevent structural hazards in pipelined designs.
Hazard Interfaces~\cite{hazardInterfaces} generalizes it further
to cover data and control hazards as well.
Both focus on higher-level notions of hazards than \issuename{}s
on high-level abstractions specialized for pipelined designs.
}

{%
\nparagraph{RTL Verification}{%
Verification techniques focus on more general specifications for RTL designs, e.g.,
those based on temporal logics~\cite{moszkowskiTemporalLogicMultilevel1985,pnueliTemporalLogicPrograms1977,camuratiFormalVerificationHardware1988,guptaFormalHardwareVerification}.
In practice, desired properties are typically specified as assertions in
source code, which are verified either through testing~\cite{Cocotb,IEEEStandardUniversal2020} or through
formal methods such as model checking~\cite{witharanaSurveyAssertionbasedHardware2022,YosysHQYosys2025,JasperFormalVerification}.
Compared with \codename{},
verification-based techniques cover more general properties, but
suffer from a long feedback loop resulting from a separate verification stage,
the extra burden of maintaining implementation-specific specifications,
and tractability issues such as state explosion.
Section~\ref{subsec:verification} compares \codename{} with verification-based
methods in more detail.%
}%
}

\section{Conclusions}

In this work,
we formalize the problem of \issuename{}s and present
\codename{}, a new hardware description language that provides timing safety by
capturing and enforcing timing requirements on shared values in timing contracts.
\codename{} ensures safe use of values which are guaranteed to remain unchanged
throughout their lifetimes.
While achieving this, it provides the capability of
cycle-level timing control and the expressiveness for describing
designs with dynamic timing characteristics.

\section*{Acknowledgments}
We thank the anonymous reviewers and
NUS KISP Lab members for their feedback, and
Yaswanth Tavva and Sai Dhawal Phaye for their help
with infrastructure setup.
This research is supported by a
Singapore Ministry of Education (MOE) Tier 2 grant MOE-T2EP20124-0007.

\bibliographystyle{ACM-Reference-Format}
\bibliography{paper,zotero}

\newpage

\appendix

\begin{figure}[t]
    \centering
    \begin{codebox}
    \begin{lstlisting}
chan ch {
  right data : (logic @res),
  left res : (logic @#1)
}

chan ch_s {
  right data : (logic @#1)
}

proc grandchild(ep : left ch_s) {
  reg cnt : logic[32];
  loop {
    set cnt := *cnt + 32'b1
  }
  loop {
    let v = if *cnt > 32'h100000 { 1'b1 } else { 1'b0 };
    send ep.data(v) >>
    cycle 1
  }
}

proc child(ep : left ch) {
  reg r : logic;
  chan ep_sl -- ep_sr : ch_s;
  spawn grandchild(ep_sl);
  loop {
    set r := ~*r >>
    let d = recv ep_sr.data >>
    send ep.data (*r & d) >>
    let _ = recv ep.res
  }
}

proc Top() {
  chan ep_sl -- ep_sr : ch;
  spawn child(ep_sl);
  loop {
    let d = recv ep_sr.data >>
    cycle 1 >>
    dprint "Value:
    cycle 1 >>
    dprint "Value should be the same
    cycle 1 >>
    send ep_sr.res (1'b1) >>
    cycle 1
  }
}
\end{lstlisting}
    \end{codebox}
    \captionof{lstlisting}{Example \codename{} code.}
    \label{lst:anvil-example}
\end{figure}

\begin{figure}[t]
    \centering
\begin{codebox}
\begin{lstlisting}[language=verilog,morekeywords={logic,always\_ff,enum,always\_comb}]
module grandchild(/* omitted */);
    logic [31:0] cnt;
    logic data_q, data_d;

    assign data_o = data_q;
    assign data_valid_o = 1'b1;
    always_comb begin
        data_d = data_q;
        if (data_ack_i) begin
            data_d = cnt > 32'h100000 ? 1'b1 : 1'b0;
        end
    end

    initial begin
        cnt <= '0;
        data_q <= 1'b0;
    end
    always_ff @(posedge clk_i) begin
        cnt <= cnt + 32'b1;
        data_q <= data_d;
    end
endmodule

module child(/* omitted */);
    /* omitted */
endmodule

module example(input logic clk_i);
    /* omitted */

    enum logic[1:0] { /* omitted */ } state_q, state_d;

    assign data_ack = state_q == RECV;
    assign res_valid = state_q == SEND;
    always_comb begin
        state_d = state_q;
        unique case (state_q)
            /* omitted */
        endcase
    end

    initial state_q <= RECV;
    always_ff @(posedge clk_i) begin
        if (state_q == ST0 || state_q == ST1) begin
            assert(data == $past(data));
        end
        state_q <= state_d;
    end
endmodule
\end{lstlisting}
\end{codebox}
    \captionof{lstlisting}{Example SystemVerilog code with assertions.}
    \label{lst:systemverilog-annotations-example}
\end{figure}

{

\section{Comparison with Verification}
\label{appx:verification}

Consider the \codename{} code in Listing~\ref{lst:anvil-example}.
The module \texttt{Top} receives data and sends back a response to the \texttt{child} module.
The data received is expected to be usable and keep unchanged in between.
The \texttt{child} module in turn communicates with grandchild to obtain the data.
\codename{}'s type system rejects this code because the data
\texttt{child} obtains from \texttt{grandchild}, namely
\texttt{d}, lives only for one cycle, but a value that depends on it (\texttt{*r \& d}) is sent to \texttt{Top} which requires it to stay alive until the response.
When fed with this code as input, the \codename{}
compiler detects this type error and
prints the following error message:
}
\begin{verbatim}
Value not live long enough in message send!
Top.anvil:29:4:
     29|     send ep.data (*r & d) >>
       |     ^^^^^^^^^^^^^^^^^^^^^
\end{verbatim}

{
Now let us consider the same design expressed in
SystemVerilog with an assertion for the same property,
provided in Listing~\ref{lst:systemverilog-annotations-example}.
Comparing the two versions leads us to the following observations:
\begin{itemize}
    \item The assertion in the SystemVerilog code is tied to the implementation.
    For example, it already requires the designer to manually and explicitly identify the states where the data is used and needs to remain unchanged.
    \codename{} does not require such manual effort.
    \item \codename{} can check the property individually for each module due to the compositional contracts specified in channel definitions.
    In the example, \codename{} can report the violation by just looking at child module alone.
    Checking the property in the SystemVerilog code requires reasoning across all three modules.
    \item Problems may arise if we attempt to apply formal verification techniques to verify the specified property
    in the SystemVerilog code.
    For example,
    bounded model checking on the SystemVerilog code using Yosys SMT-BMC and z3
    fails to detect the violation even with large depth limits.
    This is due to the large number of concrete states.
    In contrast, the \codename{} code provides abstractions that capture more of the hardware designer's intent,
    allowing the property to be checked more easily.
\end{itemize}

}

\section{Safety Analysis on Real-World Errors}

\label{appx:eval}
\label{subsec:error}

\begin{figure}[]
    \includegraphics[width=\linewidth]{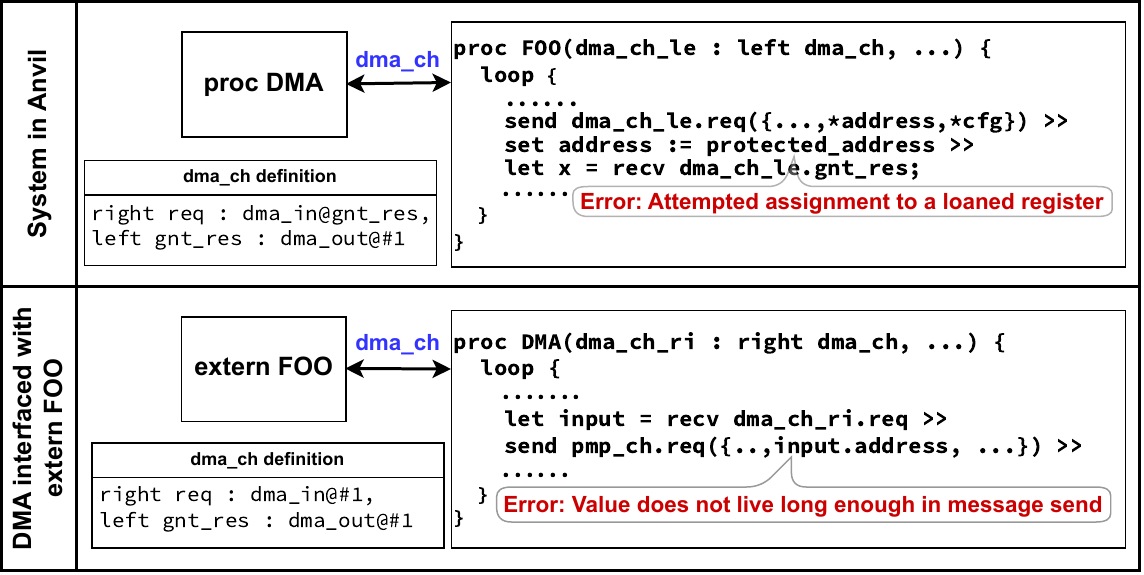}
    \caption{\codename{} can assist in preventing bugs.}
    \label{fig:bug-pick-up}
\end{figure}

\begin{table*}[t]
\caption{Summary of Issues in some open source repositories}
\label{fig:issues}
\resizebox{\textwidth}{!}{
\begin{tabular}{|l|l|l|}
\hline
\textbf{Repository} & \textbf{Issue Analysis} & \textbf{How can \codename{} help?} \\ \hline
\textbf{\begin{tabular}[c]{@{}l@{}}OpenTitan\\ (Issue~\cite{opentitan})\end{tabular}} & \begin{tabular}[c]{@{}l@{}}In OpenTitan’s entropy source module, firmware (FW) is supposed to insert\\ verified entropy data into the RNG pipeline. However, a \issuename{} \\ prevented reliable data writing and control over the SHA operation.\\ \\Solution Proposed in discussion: Add signals for FW to control the entropy\\ source state machine and a ready signal to safely write data into the pipeline.\end{tabular} & \begin{tabular}[c]{@{}l@{}}If implemented in \codename{}, FW would\\ inherently control the state machine when\\asserting data without explicit implementa\\-tion ensuring synchronization is built-in.\end{tabular} \\ \hline
\textbf{\begin{tabular}[c]{@{}l@{}}Coyote\\ (Issue~\cite{fpgasystems})\end{tabular}} & \begin{tabular}[c]{@{}l@{}}The completion queue has a 2-cycle valid signal burst instead of one cycle.\\ The issue is still open. This happens when a write request is issued on the\\ sq\_wr bus, and the cq\_wr is observed for completion. The valid signal is \\ high for 2 cycles instead of one.\\ \\ Core Issue: The timing contract was not properly implemented, though the\\ designer defined it. The timing control was deeply embedded within\\ interconnected state machines, making the bug difficult to detect even with\\ a thorough inspection.\end{tabular} & \begin{tabular}[c]{@{}l@{}}\codename{} implements the FSM for timing\\ contracts implicitly, providing synchroniza\\-tion primitives to control the state and\\ensure an error-free FSM implementation.\end{tabular} \\ \hline
\textbf{\begin{tabular}[c]{@{}l@{}}ibex\\ (Commit~\cite{ibex})\end{tabular}} & \begin{tabular}[c]{@{}l@{}}Commit Message: ``Add an instr\_valid\_id signal to completely decouple the\\ pipeline stages, hopefully, it fixes the exception controller"\\ \\ Commit Summary: Despite the pipeline being statically scheduled, the valid\\ signal was added later to enforce the timing contract only after unexpected\\ behaviour was observed.\end{tabular} & \begin{tabular}[c]{@{}l@{}}In \codename{}, even for statically scheduled \\pipelines, stage-to-stage handshakes are \\enforced implicitly, ensuring timing \\contracts are upheld even if the\\ schedule isn’t strictly adhered to.\end{tabular} \\ \hline
\textbf{\begin{tabular}[c]{@{}l@{}}snax-cluster\\ (Commit~\cite{snaxcluster})\end{tabular}} & \begin{tabular}[c]{@{}l@{}}Commit changes\\ \texttt{assign} \texttt{a\_ready\_o} = \texttt{acc\_ready\_i} \texttt{\&\&} \texttt{c\_ready\_i} \texttt{\&\&} (\texttt{a\_valid\_i} \texttt{\&\&} \texttt{b\_valid\_i});\\ \texttt{assign} \texttt{b\_ready\_o} = \texttt{acc\_ready\_i} \texttt{\&\&} \texttt{c\_ready\_i} \texttt{\&\&} (\texttt{a\_valid\_i} \texttt{\&\&} \texttt{b\_valid\_i});\\ \\Commit Summary: Fixes the implementation of the timing contract on the ALU\\ interface by adding the missing valid signal in the handshake.\end{tabular} & \begin{tabular}[c]{@{}l@{}}\codename{} implicitly handles handshake impl\\-ementation for interfacing signals,\\ ensuring the enforcement of timing\\contracts.\end{tabular} \\ \hline
\textbf{\begin{tabular}[c]{@{}l@{}}core2axi\\ (Commit~\cite{core2axi})\end{tabular}} & \begin{tabular}[c]{@{}l@{}}Commit changes: \texttt{w\_valid\_o = 1'b1;}\\ \\ Commit Summary: Ensure compliance with the timing contract by asserting\\ the missing valid signal when sending a new write request on the bus.\end{tabular} & \begin{tabular}[c]{@{}l@{}}In \codename{}, the assertion of valid signals and\\ synchronization is handled implicitly\\whenever a message is sent\end{tabular} \\ \hline
\end{tabular}
}
\end{table*}

We were motivated to design \codename{} by our own frustrating experience implementing
an experimental CPU architecture.
The frequent \issuename{} we encountered during development
required significant debugging effort.
We demonstrate how \codename{} can help designers address the following challenges with minimal effort:

\begin{enumerate}
\item Enforcing concrete timing contracts
\item Challenges in implementing timing contracts
\end{enumerate}

\nparagraph{Case 1: Enforcing Concrete Timing Contracts}{}
The vulnerability class highlighted in CWE-1298~\cite{MITRECWE1298} illustrates a hardware bug from HACK@DAC'21.
This bug arose from a missing timing contract in the DMA module of the OpenPiton SoC.
The module was intended to verify access to protected memory using specific address and configuration signals. However, it assumed these inputs would remain stable during processing without any mechanism to enforce this assumption. This created a timing vulnerability across module interactions.

If designed in \codename{}, the DMA channel definition would explicitly require that input signals remain stable until the request is completed, as shown in Figure~\ref{fig:bug-pick-up}.
\codename{} would enforce this stability requirement, ensuring that only compatible modules interact without introducing timing risks. When the DMA module interfaces with non-\codename{} modules, \codename{} imposes a one-clock-cycle lifetime on external signals. If the DMA implementation does not follow the contract, \codename{} triggers an error: ``Value does not live long enough\ldots,'' implying the need to register the signal immediately.

Similarly, designers using custom test benches with open-source hardware often struggle to follow strict timing contracts. This is particularly challenging when there is no mechanism to enforce timing contracts. For instance, in this GitHub issue~\cite {etherveri}, the designer observed unexpected behaviour during simulation while integrating a Verilog-based Ethernet interface into their module.
This Ethernet module required a complex timing contract to be enforced on the interfacing module for proper operation. However, without a language that enforces this contract, the designer struggled to explicitly meet these timing requirements and manage synchronization.

\nparagraph{Case 2: Challenges in Implementing Timing Contracts}{}
Designers often face challenges in implementing synchronization
primitives and dynamic timing contracts, even when they intend to define them clearly.
This difficulty is evident in various open-source project commit histories and issue trackers.
For example, in Table~\ref{fig:issues}, we highlight a few instances
from GitHub that showcase how designers have struggled with these aspects. Our analysis demonstrates that \codename{} could have prevented these issues or helped catch the bugs before compilation.

Even when contracts are explicitly defined, the instructions
for compliance can be ambiguous.
A case in point is the documentation for CV-X-IF, where one
issue~\cite{corevdoc} reveals the complications
involved in adhering to the timing contract.
Another issue~\cite{corevdo} illustrates that the complexity of
a static schedule necessitated additional notes to
clarify the implementation guidelines for the interfacing module.

In contrast, \codename{} simplifies the implementation of synchronization and finite state machines (FSM) that handle timing contracts. Designers only need to define the contract within the corresponding channel, which can utilize dynamic message-passing events. The synchronization primitives (handshakes) are implemented implicitly and efficiently, ensuring no clock cycle overhead. Additionally, the wait construct allows designers to express the dynamic times required to process a state. In ambiguous process descriptions, \codename{} flags the description to make necessary changes to guarantee runtime safety statically.

\onecolumn
\section{Formalization Details}
\label{appx:details}

\subsection{Abstract Syntax}

\begin{figure}[t]
    \small
        \begin{equation*}
        \text{message definitions} \quad M ::= \{ \pi.m : p, \cdots \}
        \end{equation*}
        \begin{equation*}
        \text{message set} \quad \Sigma ::= \{ \pi.m, \cdots \}
        \end{equation*}
        \begin{equation*}
        \text{composition} \quad \kappa ::= t \mid \kappa \parallel_\Sigma \kappa
        \end{equation*}
        \begin{equation*}
        \text{program} \quad \program ::= (\texttt{loop} \{ t \}, M) \mid \program \parallel_\Sigma \program
        \end{equation*}
    \caption{\codename{} abstract syntax}
    \label{fig:abstract-syntax}
\end{figure}

For convenience of formal reasoning, we also define an abstract syntax of
\codename{} programs, shown in Figure~\ref{fig:abstract-syntax},
allowing us to discuss parallel composition in a style
similar to communicating sequential processes (CSP)~\cite{hoareCommunicatingSequentialProcesses1978}.
The $\parallel_\Sigma$ notation represents parallel composition with the two sides
communicating through messages specified in the set $\Sigma$.
$M$ maps each message to the associated duration requirement.

\subsection{Semantics}

\nparagraph{Execution log}{
An execution log is simply a sequence $\exlog = \langle \alpha_0, \cdots, \alpha_k \rangle$,
where $\alpha_i$ is represents the set of operations performed during cycle $i$.
Operations can be one of the following ---
\begin{enumerate*}
    \item \textbf{ValCreate} representing the creation of a new value that depends on a set of registers and existing
    values,
    \item \textbf{ValUse}, representing the use of a value,
    \item \textbf{RegMut}, denoting mutation of a register,
    \item \textbf{ValSend}, for sending of a value through a message, and
    \item \textbf{ValRecv}, denoting the receipt of a value through a message.
\end{enumerate*}
}
Following this, we define the set of execution logs corresponding to a term, compositions, and finally programs.
To capture the non-determinism of message passing and branching in an execution log
of a term, we delay each send and receive operation by any non-negative
number of cycles and allow each branching term to take either branch.
Execution logs of compositions are obtained by combining two execution logs,
with the requirement that any send and receive operations for messages in $\Sigma$
must match and align in pairs, and each pair must use the same value identifier.
In the combined execution log, the matching send and receive operations are eliminated. This reflects that they have now become internal details, no longer affecting the semantics of the composition.
For programs, we take into consideration the looping semantics of each looping thread.
We achieve this by mapping a program to a set of compositions, where each composition is
obtained by appending $t$ in each looping thread $\texttt{loop}\{ t \}$ arbitrarily many times.
Any execution log of any such composition is an execution log of the program.
The semantics of those constructs is then defined by their sets of execution logs,
which captures all their possible behaviours.

\begin{definition}[Execution log]
    An execution log consists of a sequence of sets $\exlog = \langle \alpha_0, \alpha_1, \cdots, \alpha_k \rangle$.
    The finite set $\alpha_i$ contains the actions in the $i$-th cycle, each of the following form:
    \begin{itemize}
        \item $\valuecreate{v}{\{r_1, r_2, \cdots, r_m\}}{\{v_1, v_2, \cdots, v_n\}}$ (creating a value with name $v$ that
        depends on registers $r_1, r_2, \cdots, r_m$ and values
        $v_1, v_2, \cdots, v_n$)
        \item $\valueuse{v}$ (using the value identified by $v$)
        \item $\regmutate{r}$ (mutating the register identified by $r$)
        \item $\valuesend{\pi.m}{v}{p}$ (send a value with name $v$ through message $\pi.m$ with duration $p$)
        \item $\valuerecv{\pi.m}{v}{p}$ (receive a value with name $v$ through message $\pi.m$ with
        duration $p$)
    \end{itemize}
\end{definition}

\begin{definition}[Local execution log]
A log $\exlog$ is a local execution log of a term $t$ if
$\ejudge{\Gamma}{I}{M}{\evalto{t}{\exlog}{v}{S}}$, which is defined
by the following inference rules.
\begin{equation*}
\inferrulena{}{
\ejudge{\Gamma}{\{v\}}{M}{\evalto{\texttt{cycle } \#k}{(\emptyset^{k + 1} \circ \langle\{\valuecreate{v}{\emptyset}{\emptyset}\}\rangle)}{v}{\emptyset}}
}{E-Cycle}
\end{equation*}

\begin{equation*}
\inferrulena{}{
\ejudge{\Gamma}{\{v\}}{M}{\evalto{n}{\langle\{\valuecreate{v}{\emptyset}{\emptyset}\}\rangle}{v}{\emptyset}}
}{E-Literal}
\end{equation*}

\begin{equation*}
\inferrulena{
\ejudge{\Gamma}{I_1}{M}{
\evalto{t_1}{\exlog_1}{v_1}{S_1}
} \quad
\ejudge{\Gamma}{I_2}{M}{
\evalto{t_2}{\exlog_2}{v_2}{S_2}
} \\
I_1 \cap I_2 = \emptyset
}{
\ejudge{\text{shift}(\Gamma, |\exlog_1| - 1)}{(I_1 \cup I_2)}{M}{
\evalto{t_1 \texttt{ => } t_2}{(\exlog_1 \circ \exlog_2)}{v_2}{S_2}
}
}{E-Wait}
\end{equation*}

\begin{equation*}
\inferrulena{
\ejudge{\Gamma}{I_1}{M}{
\evalto{t_1}{\exlog_1}{v_1}{S_1}
}
\quad
\ejudge{\Gamma, x : (|\exlog_1| - 1, v_1)}{I_2}{M}{
\evalto{t_2}{\exlog_2}{v_2}{S_2}
} \\
}{
\ejudge{\Gamma}{(I_1 \cup I_2)}{M}{
\evalto{\texttt{let } x = t_1 \texttt{ in } t_2}{(\exlog_1 \uplus \exlog_2)}{v_2}{S_2}
} \\
I_1 \cap I_2 = \emptyset
}{E-Let}
\end{equation*}

\begin{equation*}
\inferrulena{
\Gamma(x) = (k, v)
}{
\ejudge{\Gamma}{\emptyset}{M}{
\evalto{x}{\emptyset^{k + 1}}{v}{S}
}
}{E-Ref}
\end{equation*}

\begin{equation*}
\inferrulena{
    \ejudge{\Gamma}{I}{M}{
        \evalto{t}{\exlog}{v}{S}
    } \quad
    v^\prime \not\in I
}{
    \ejudge{\Gamma}{I \cup \{v^\prime\}}{M}{
         \evalto{r := t}{\exlog \uplus \langle\{\valueuse{v}, \regmutate{r}\}, \{\valuecreate{v^\prime}{\emptyset}{\emptyset}\}\rangle}{v^\prime}{\emptyset}
    }
}{E-RegAssign}
\end{equation*}

\begin{equation*}
\inferrulena{
    \ejudge{\Gamma}{I}{M}{
        \evalto{t}{\exlog}{v}{S}
    } \quad
    v^\prime \not\in I \quad
    k \in \mathbb{N}
}{
    \ejudge{\Gamma}{(I \cup \{v^\prime\})}{M}{
        \evalto{\texttt{send } \pi.m (t)}{\emptyset^{k + 1} \circ \langle\{\valuesend{\pi.m}{v}{M(\pi.m)},
        \valuecreate{v^\prime}{\emptyset}{\emptyset}\} \rangle}{v^\prime}{\emptyset}
    }
}{E-Send}
\end{equation*}

\begin{equation*}
\inferrulena{
    k \in \mathbb{N},
    u \neq v
}{
    \ejudge{\Gamma}{(\{v, u\})}{M}{
        \evalto{\texttt{recv } \pi.m}{\emptyset^k \circ \\ \langle\{\valuerecv{\pi.m}{v}{M(\pi.m)},
        \valuecreate{u}{\emptyset}{\{v\}}\}\rangle}{u}{\emptyset}
    }
}{E-Recv}
\end{equation*}

\begin{equation*}
\inferrulena{
    \ejudge{\Gamma}{I_1}{M}{
    \evalto{t_1}{\exlog_1}{v_1}{S_1}} \\
    \ejudge{\Gamma}{I_2}{M}{
    \evalto{t_2}{\exlog_2}{v_2}{S_2}} \\
    \ejudge{\Gamma}{I_3}{M}{
    \evalto{t_3}{\exlog_3}{v_3}{S_3}} \\
    I_1 \cap (I_2 \cup I_3) = \emptyset \quad I_2 \cap I_3 = \emptyset
}{
    \ejudge{\Gamma}{(I_1 \cup I_2 \cup I_3)}{M}{
        \evalto{\texttt{if } t_1 \texttt{ then } t_2 \texttt{ else } t_3}{\exlog_1 \uplus \exlog_2 \uplus \langle \{ \valueuse{v_1} \}\rangle }{v_2}{S_2}
    }
}{E-IfThen}
\end{equation*}

\begin{equation*}
\inferrulena{
    \ejudge{\Gamma}{I_1}{M}{
    \evalto{t_1}{\exlog_1}{v_1}{S_1}} \\
    \ejudge{\Gamma}{I_2}{M}{
    \evalto{t_2}{\exlog_2}{v_2}{S_2}} \\
    \ejudge{\Gamma}{I_3}{M}{
    \evalto{t_3}{\exlog_3}{v_3}{S_3}} \\
    I_1 \cap (I_2 \cup I_3) = \emptyset \quad I_2 \cap I_3 = \emptyset
}{
    \ejudge{\Gamma}{(I_1 \cup I_2 \cup I_3)}{M}{
        \evalto{\texttt{if } t_1 \texttt{ then } t_2 \texttt{ else } t_3}{\exlog_1 \uplus \exlog_3 \uplus \langle \{ \valueuse{v_1} \}\rangle }{v_3}{S_3}
    }
}{E-IfElse}
\end{equation*}

\begin{equation*}
\inferrulena{
}{
    \ejudge{\emptyset}{\{v\}}{M}{
        \evalto{*r}{\langle\{\valuecreate{v}{\{r\}}{\emptyset}\} \rangle}{v}{\{r\}}
    }
}{E-RegEval}
\end{equation*}

\begin{equation*}
\inferrulena{
}{
\ejudge{\emptyset}{\{v\}}{M}{
    \evalto{\texttt{ready}(\pi.m)}{\langle \{\valuecreate{v}{\emptyset}{\emptyset} \} \rangle}{v}{v}
}
}{E-Ready}
\end{equation*}

Where $\langle \alpha_0, \alpha_1, \cdots, \alpha_{k} \rangle \circ \langle \beta_0, \beta_1, \cdots, \beta_{l} \rangle = \langle \alpha_0, \alpha_1, \cdots, (\alpha_k \cup \beta_0), \beta_1, \cdots,
\beta_l \rangle$.

The merge operator $\uplus$ is defined as (without loss of generality, assuming $k \leq l$):
$\langle \alpha_0, \alpha_1, \cdots, \alpha_k \rangle \uplus \langle \beta_0, \beta_1, \cdots, \beta_l \rangle
= \langle \alpha_0 \cup \beta_0, \alpha_1 \cup \beta_1, \cdots, \alpha_k \cup \beta_k, \beta_{k + 1}, \cdots, \beta_l \rangle$.

$\alpha^k = \langle \alpha_0, \cdots, \alpha_{k - 1} \rangle$ where for all $i = 0, 1, \cdots, k - 1$,
$\alpha_i = \alpha$.

The function $\text{shift}(\Gamma, k)$ shifts all delays in $\Gamma$ by $k$ cycles. Formally,
\begin{align*}
\text{shift}(\emptyset, k) = \emptyset \\
\text{shift}((\Gamma, x : (k^\prime, v)), k) =
\text{shift}\left (\Gamma, k), x : (\max (0, k^\prime - k), v \right )
\end{align*}

\end{definition}

\begin{definition}[Compositional execution log]
$\exlog$ is an execution log of a $\kappa$ if:
\begin{itemize}
    \item $\kappa = t$ and $\exlog$ is a prefix of an execution log of $t$
    \item $\kappa = \kappa_1 \parallel_\Sigma \kappa_2$, $\exlog_1, \exlog_2$ are
    execution logs of $\kappa_1$ and $\kappa_2$ respectively, and
let $\exlog_1 = \langle \alpha_0, \cdots, \alpha_m \rangle,
\exlog_2 = \langle \beta_0, \cdots, \beta_m \rangle$, the following
holds:
    \begin{itemize}
        \item For all $\pi.m \in \Sigma$, $0 \leq i \leq m$, $\valuesend{\pi.m}{v}{p} \in \alpha_i$ if and only
        if $\valuerecv{\pi.m}{v}{p} \in \beta_i$, and
        $\valuerecv{\pi.m}{v}{p} \in \alpha_i$ if and only if $\valuesend{\pi.m}{v}{p} \in \beta_i$.
        \item $\exlog = \langle \gamma_0, \cdots, \gamma_m \rangle, \gamma_i =
        \alpha_i \cup \beta_i - \{\valuesend{\pi.m}{v}{p} \mid \pi.m \in \Sigma\} - \{\valuerecv{\pi.m}{v}{p} \mid \pi.m \in \Sigma\}$.
    \end{itemize}
\end{itemize}
\end{definition}

\newcommand{\termcomp}[3]{#1 \parallel_{#2} #3}

\begin{definition}[Concretization]
    A composition $\kappa$ is a concretization of program $\program$, written
    $\program \leadsto \kappa$, by the following inference rules:
    \begin{equation*}
        \inferrulena{}{(\texttt{loop} \{ t\}, M) \leadsto t}{C-Base}
    \end{equation*}
    \begin{equation*}
        \inferrulena{(\texttt{loop} \{ t\}, M) \leadsto t^\prime}{(\texttt{loop} \{ t\}, M) \leadsto t^\prime \texttt{ => } t}{C-Extend}
    \end{equation*}
    \begin{equation*}
        \inferrulena{\program_1 \leadsto \kappa_1 \quad \program_2 \leadsto \kappa_2}{
            \program_1 \parallel_\Sigma \program_2 \leadsto \kappa_1 \parallel_\Sigma \kappa_2
        }{C-Compose}
    \end{equation*}
\end{definition}

\begin{definition}[Program execution log]
   $\exlog$ is an execution log of program $\program$ if
   there exists composition $\kappa$ such that $\program \leadsto \kappa$ and
   $\exlog$ is an execution log of $\kappa$.
\end{definition}

\subsection{Type System}\label{subsec:typeSysformal}

\nparagraph{Event graph}{
The type system of \codename{} is based on the \emph{event graph}.
An event graph, denoted $G = (V, E)$,
is a directed acyclic graph that describes
the time ordering among events in an \codename{} process.
Each node (i.e., event)
is labelled to indicate how its corresponding starting time relates
to those of its direct predecessors.
Types in \codename{} reference the event graph as part of the typing environment
to convey timing constraints.
We choose this strategy
because the timing constraints associated with a term
are not always local.
Take the example of \prog{send ch.m1 (x) => recv ch.m2}, where
\texttt{ch.m1} specifies a duration of \texttt{ch.m2}.
It is necessary to be aware of the first \texttt{ch.m2} event that occurs after \texttt{ch.m1}. This event does not appear in the expression \prog{send ch.m1 (x)} itself, but rather in the surrounding context in which \prog{send ch.m1 (x)} appears, to ensure that \texttt{x} lives long enough.

We choose the event graph as
it is a simple structure that captures all the necessary information to reason about such timing constraints.
As a shorthand, we use
the notation $e_1 \to e_2 \in G$ to say that $G$ contains an edge
from event $e_1$ to event $e_2$.
We use $G(e_2)$ to denote
$\graphpred{\omega}{\{e_1 \mid e_1 \to e_2 \in G\}}$, which consists of the operation label
$\omega$ of $e_2$ as well as the set of all its direct predecessors.
}

\nparagraph{Types}{
Intuitively, a type encodes a lifetime
by referencing the event graph and is a pair:
\begin{align*}
    T & ::= (e_l, S_d),
\end{align*}
where $e_l$ is an event graph node that encodes the start time,
and $S_d$ is a set of event patterns
$\timepattern{e_d}{p}$, the earliest match of which defines
the end time.
An empty $S_d$ indicates that the lifetime is eternal.
Each time pointer specifier is a pair of event identifier $e_d$ and
duration $p$, which implies the first time $p$ is matched (the specified number
of cycles have elapsed or a specified message is sent or received)
after $e_d$ is reached.
}

\nparagraph{Typing Rules}{
A typing judgment is of the form
\begin{equation*}
    \tjudge{\Gamma}{G}{R}{M}{e_c}{t : T}.
\end{equation*}
The typing environment consists of $\Gamma$ which maps each let-binding to its type,
the event graph $G$ introduced above, $R$ which maps a register to its loan time,
$M$ which maps a message specifier (an endpoint and a message identifier, of the form $\pi.m$)
to the duration that specifies its lifetime requirement,
$C$ which is a set of identifiers associated with all branch conditions that have appeared,
and $e_c$ which references a node in $G$ as an abstract specifier of the time at which
$t$ is to be evaluated.

The typing rules use the $\leq_G$ and $<_G$ relations to apply timing constraints.
Their complete and formal definitions are available in Section~\ref{appx:details}.
Intuitively, $a \leq_G b$ if the time specified by $a$
is always no later than that by $b$ in the event graph $G$,
and $a <_G b$ if the time specified by $a$
is always strictly before that by $b$ in $G$.
Here $a$ and $b$ can be nodes or timing patterns in $G$.
In our implementation, we use sound approximations of $\leq_G$ and $<_G$.
}

\begin{equation*}
\inferrulena{
    \tjudgefull{\Gamma}{G}{R}{M}{C}{e_c}{t : T} \quad
}{
    \tjudgefull{\Gamma, x : T^\prime}{G}{R}{M}{C}{e_c}{t : T}
}{T-Weaken}
\end{equation*}

\begin{equation*}
\inferrulena{
    G(e_l) = \graphpred{\# k}{\{e_c\}}
}{
    \tjudgefull{\emptyset}{G}{R}{M}{\emptyset}{e_c}{\texttt{cycle } k : (e_l, \emptyset)}
}{T-Cycle}
\end{equation*}

\begin{equation*}
\inferrulena{
    \tjudgefull{\Gamma}{G}{R}{M}{C_1}{e_c}{t_1 : (e_l, S_d)} \\
    \tjudgefull{\Gamma}{G}{R}{M}{C_2}{e_l}{t_2 : T_2} \quad C_1 \cap C_2 = \emptyset
}{\tjudgefull{\Gamma}{G}{R}{M}{C_1 \cup C_2}{e_c}{t_1 \texttt{ => } t_2 : T_2}}{T-Wait}
\end{equation*}

\begin{equation*}
\inferrulena{
    M(\pi.m) = p \quad G(e_l) = \graphpred{\pi.m}{\{e_c\}}
}{
    \tjudgefull{\emptyset}{G}{R}{M}{\emptyset}{e_c}{\texttt{recv } m : (e_l, \{\timepattern{e_l}{p}\})}
}{T-Recv}
\end{equation*}

\begin{equation*}
\inferrulena{
    x : (e_l, S_d) \in \Gamma \quad G(e^\prime_l) = \graphpred{\#0}{\{e_c, e_l\}}
}{
    \tjudgefull{\Gamma}{G}{R}{M}{\emptyset}{e_c}{x : (e^\prime_l, S_d)}
}{T-Ref}
\end{equation*}

\begin{equation*}
\inferrulena{
    \tjudgefull{\Gamma}{G}{R}{M}{C}{e_c}{t : (e_l, S_d)} \\
    G(e^\prime_l) = \graphpred{\pi.m}{\{e_c\}} \\
    \nolater{G}{e_l}{e_c} \quad
    \nolater{G}{\timepattern{e^\prime_l}{M(\pi.m)}}{S_d}
}{
    \tjudgefull{\Gamma}{G}{R}{M}{C}{e_c}{\texttt{send } \pi.m (t) : (e^\prime_l, \emptyset)}
}{T-Send}
\end{equation*}

\begin{equation*}
\inferrulena{
    \tjudgefull{\Gamma}{G}{R}{M}{C_1}{e_c}{t_1 : (e_1, S_1)} \\
    \tjudgefull{\Gamma}{G}{R}{M}{C_2}{e_c}{t_2 : (e_2, S_2)} \\
    G(e^\prime_l) = \graphpred{\#0}{\{e_1, e_2\}} \quad C_1 \cap C_2 = \emptyset
}{
    \tjudgefull{\Gamma}{G}{R}{M}{C_1 \cup C_2}{e_c}{t_1 \star t_2 : (e^\prime_l, S_1 \cup S_2)}
}{T-BinOp}
\end{equation*}

\begin{equation*}
\inferrulena{
    \tjudge{\Gamma}{G}{R}{M}{e_c}{t : (e_l, S_d)} \\
    \forall (e, S) \in R(r): \earlierthan{G}{e_c}{e} \lor
    \nolater{G}{S}{e_c} \\
    \nolater{G}{e_l}{e_c} \quad \nolater{G}{\timepattern{e_c}{\#1}}{S_d} \quad
    G(e^\prime_l) = \graphpred{\#1}{\{e_c\}}
}{
    \tjudge{\Gamma}{G}{R}{M}{e_c}{r := t : (e^\prime_l, \emptyset)}
}{T-RegAssign}
\end{equation*}

\begin{equation*}
\inferrulena{
    \exists (e, S) \in R(r): \nolater{G}{e}{e_c} \land \nolater{G}{e_c}{S_d} \land
    \nolater{G}{S_d}{S}
}{
    \tjudgefull{\emptyset}{G}{R}{M}{\emptyset}{e_c}{*r : (e_c, S_d)}
}{T-RegEval}
\end{equation*}

\begin{equation*}
\inferrulena{
    \tjudgefull{\Gamma}{G}{R}{M}{C_1}{e_c}{t_1 : (e_1, S_1)} \\
    \tjudgefull{\Gamma}{G}{R}{M}{C_2}{e^\prime_c}{t_2 : (e_2, S_2)} \\
    \tjudgefull{\Gamma}{G}{R}{M}{C_3}{e^{\prime\prime}_c}{t_3 : (e_3, S_3)} \\
    \nolater{G}{e_1}{e_c} \land \nolater{G}{e_c}{S_1} \\
    c \not\in C_1 \cup C_2 \cup C_3 \quad C_1 \cap (C_2 \cup C_3) = \empty \quad C_2 \cap C_3 = \emptyset \\
    \quad G(e^\prime_c) = G(e^{\prime\prime}_c) = \graphpred{\&c}{\{e_c\}} \quad
    e^\prime_c \neq e^{\prime\prime}_c \\
    G(e^\prime_l) = \graphpred{\oplus}{\{e_2, e_3\}}
}{
    \tjudgefull{\Gamma}{G}{R}{M}{C_1 \cup C_2 \cup C_3 \cup \{c\}}{e_c}{\texttt{if } t_1 \texttt{ then } t_2 \texttt{ else } t_3 : (e^\prime_l, S_1 \cup S_2 \cup S_3)}
}{T-Cond}
\end{equation*}

\begin{equation*}
\inferrulena{
    \tjudgefull{\Gamma}{G}{R}{M}{C_1}{e_c}{t_1 : (e_1, S_1)} \\
    \tjudgefull{\Gamma}{G}{R}{M}{C_2}{e_c}{t_2 : (e_2, S_2)} \\
    G(e^\prime_l) = \graphpred{\#0}{\{e_1, e_2\}} \quad C_1 \cap C_2 = \emptyset
}{
    \tjudgefull{\Gamma}{G}{R}{M}{C_1 \cup C_2}{e_c}{t_1; t_2 : (e^\prime_l, S_2)} \\
}{T-Join}
\end{equation*}

\begin{equation*}
\inferrulena{
    (\pi.m, p) \in M
}{
    \tjudgefull{\emptyset}{G}{R}{M}{\emptyset}{e_c}{\texttt{ready}(\pi.m) : (e_c, \{\timepattern{e_c}{\#1}\})} \\
}{T-Ready}
\end{equation*}

\nparagraph{Well-typedness}{
We define well-typed terms, processes, and programs based on the above.

\begin{definition}[Well-typed \codename{} term]
An \codename{} term $t$ is well-typed under the context $M$ if
there exist $G$, $R$, $e_0$, $C$, and $T$ such that $G(e_0) = \graphpred{0}{\emptyset}$ and
$\tjudgefull{\emptyset}{G}{R}{M}{C}{e_0}{t : T}$.
\end{definition}

\begin{definition}[Well-typed \codename{} process]
Under the context $M$, we say
a process loop $\texttt{loop} \{ t \}$ is well-typed if
the term $t \Rightarrow t$ is well-typed under $M$.
\end{definition}

\begin{definition}[Well-typed \codename{} program]
A program $\program$ is well-typed if
\begin{itemize}
    \item $\program = (\texttt{loop} \{ t \}, M)$ and $\texttt{loop} \{ t \}$ is well-typed under $M$.
    \item $\program = \program_1 \parallel_\Sigma \program_2$, and $\Sigma = M_{\program_1} \cap M_{\program_2}$,
    where $M_{\program_i}$ is the union of all $M$s that appear in $\program_i$.
\end{itemize}
\end{definition}
}

\subsubsection{Auxiliary Definitions}

We define $\leq_G$ and $<_G$ that appear in the typing rules.

\begin{definition}[Timestamp]
    A function $\tau_G : V \to \mathbb{N}$ is a timestamp function of event graph
    $G = (V, E)$ if for all $e \in V$:
    \begin{itemize}
    \item If $G(e) = \graphpred{0}{S}$, then $\timestamp{G}{e} = 0$.
    \item If $G(e) = \graphpred{\#k}{S}$, then $\timestamp{G}{e} = \max_{e^\prime \in S}
    \left ( \timestamp{G}{e^\prime} + k \right )$.
    \item If $G(e) = \graphpred{\pi.m}{S}$, then $\timestamp{G}{e} \geq \max_{e^\prime \in S}
    \timestamp{G}{e^\prime}$
    \item If $G(e) = \graphpred{\&c}{S} \land \timestamp{G}{e} = \max_{e^\prime \in S}
    \timestamp{G}{e^\prime}$, then $\forall e^\prime \in V : (e^\prime \neq e \land G(e) = \graphpred{\&c}{S}) \to \timestamp{G}{e^\prime} = \infty$
    \item If $G(e) = \graphpred{\oplus}{S}$, then $\timestamp{G}{e} = \min_{e^\prime \in S}
    \timestamp{G}{e^\prime}$.
    \end{itemize}
\end{definition}

It is obvious that for any event graph $G$, at least one
timestamp function exists.
We now extend this definition of timestamps to event patterns.
\begin{definition}[Event pattern timestamp]
Let $G$ be an event graph and $\tau_G$ be a timestamp function of $G$.
We define $\timepattern{e}{p}$:
\begin{itemize}
    \item $\timestamp{G}{\timepattern{e}{\#k}} = \timestamp{G}{e} + k$
    \item $\timestamp{G}{\timepattern{e}{\pi.m}} = \min_{G(e^\prime) = \graphpred{\pi.m}{S},
    \timestamp{G}{e} < \timestamp{G}{e^\prime}} \timestamp{G}{e^\prime}$ (or $\infty$
    if no such $e^\prime$ can be found).
\end{itemize}
\end{definition}

\begin{definition}[$\leq_G$ and $<_G$]
Let $G$ be an event graph.
We say $\timepattern{e_1}{p_1} \leq_G \timepattern{e_2}{p_2}$ if
for all timestamp functions $\tau_G$ of $G$, it holds that
$\timestamp{G}{\timepattern{e_1}{p_1}} \leq \timestamp{G}{\timepattern{e_2}{p_2}}$.
Similarly, we say
$\timepattern{e_1}{p_1} <_G \timepattern{e_2}{p_2}$ if
for all timestamp functions $\tau_G$ of $G$, it holds that
$\timestamp{G}{\timepattern{e_1}{p_1}} < \timestamp{G}{\timepattern{e_2}{p_2}}$.
\end{definition}

It is easy to prove the following two lemmas.

\begin{lemma}
    If $(e_1 \to e_2) \in G$, then $e_1 \leq_G e_2$.
\end{lemma}

\begin{lemma}
$S \cup S^\prime \leq_G S$.
\end{lemma}

\subsection{Safety}

\begin{definition}[Register dependency set]
We define that the value $v$ has the register dependency set $D$ in the execution log
$\exlog$, written $\regdep{\exlog}{v}{D}$, by the following inference rules:

\begin{equation*}
    \inferrulena{
    }{
        \regdep{\langle \rangle}{v}{\bot}
    }{R-Base}
\end{equation*}

\begin{equation*}
    \inferrulena{
        \regdep{\exlog}{v}{D}
    }{
        \regdep{\exlog \cdot \langle \emptyset \rangle}{v}{D}
    }{R-Empty}
\end{equation*}

\begin{equation*}
    \inferrulena{
        \regdep{\exlog \cdot \langle \alpha_i \rangle}{v}{D} \\
        o \not\in \{\valuecreate{v}{S_r}{S_v} \mid S_r \in 2^{\text{RegId}}, S_v \in 2^{\text{ValId}}\}
    }{
        \regdep{\exlog \cdot \langle \alpha_i \cup \{o\} \rangle}{v}{D}
    }{R-NonCreate}
\end{equation*}

\begin{equation*}
    \inferrulena{
        \regdep{\exlog \cdot \langle \alpha_i \rangle}{v_1}{D_1} \quad D_1 \neq \bot \\
        \vdots \\
        \regdep{\exlog \cdot \langle \alpha_i \rangle}{v_k}{D_k} \quad D_k \neq \bot \\
    }{
        \regdep{\exlog \cdot \langle \alpha_i \cup \{ \valuecreate{v}{S_r}{\{v_1, \cdots, v_k\}} \} \rangle }{v}{
            S_r \cup D_1 \cup \cdots \cup D_k
        }
    }{R-Create}
\end{equation*}

Note: $\cdot$ is the normal concatenation operator.

Other auxiliary definitions, assuming
$\exlog = \langle \alpha_0, \alpha_1, \cdots, \alpha_k \rangle$,
\begin{itemize}
    \item $\useset{\exlog}{v} = \{i \mid \valueuse{v} \in \alpha_i \lor \valuecreate{v}{S_r}{S_v} \in \alpha_i \lor \valuerecv{\pi.m}{v}{p} \in \alpha_i \lor \valuesend{\pi.m}{v}{p} \in \alpha_i\}$
    \item $\mutset{\exlog}{D} = \{i \mid r \in D \land \regmutate{r} \in \alpha_i\}$
    \item $\ltrecv{\exlog}{v} = \bigcap_{u \in \depset{\exlog}{v}, \valuerecv{\pi.m}{u}{p} \in \alpha_i}
    \ltfun{\exlog}{i}{p}$
    \item $\ltsend{\exlog}{v} = \bigcup_{u \in \deriveset{\exlog}{v}, \valuesend{\pi.m}{u}{p} \in \alpha_i}
    \ltfun{\exlog}{i}{p}$
    \item $\ltfun{\exlog}{i}{\pi.m} = [i, w)$ where $w$ is the lowest $j \geq i$, such that $\valuesend{\pi.m}{v}{p} \in \alpha_j$ or $\valuerecv{\pi.m}{v}{p} \in \alpha_j$
    \item $\ltfun{\exlog}{i}{\#l} = [i, i + l)$
\end{itemize}

\end{definition}

\nparagraph{Defining safety}{
We first define when an execution log should be deemed safe.
This notion, then, can be naturally lifted to define the safety of a term, composition of terms and of an entire \codename{} program.

\begin{definition}[Safety of execution log]
An execution log $\exlog$ is safe if
for every value $v$, there exists an interval $[a, b]$ such that
$\useset{\exlog}{v} \cup \ltsend{\exlog}{v} \subseteq [a, b] \subseteq \ltrecv{\exlog}{v}$, and
for $D$ such that $\regdep{\exlog}{v}{D}$,
$\mutset{\exlog}{D} \cap [a, b) = \emptyset$.
\end{definition}
$\useset{\exlog}{v}$ includes all time points (cycle numbers) at which
the value $v$ is used, $\ltsend{\exlog}{v}$ captures when $v$ needs to be live
as required by all send operations that involve $v$ or other values that depend on it,
$\ltrecv{\exlog}{v}$ captures when $v$ is guaranteed to be live through received messages
from the environment, $\regdep{\exlog}{v}{D}$ states that $v$ directly or indirectly depends
on the set of registers $D$, and $\mutset{\exlog}{D}$ captures when any register in $D$ is mutated.
Intuitively, the safety definition above states that all uses of a value \( v \) and the lifetime promised to the environment should fall within a continuous time window. During this time window, values received from the environment through \textit{receive} are live, and no register that $v$ depends on is mutated.

Since the set of all execution logs of a term, composition, or program captures all
its possible run-time timing behaviours, we define safety for those constructs as follows.
\begin{definition}[Term, composition, and program safety]
A term, composition, or program is safe if all its execution logs are safe.
\end{definition}
}

\nparagraph{Safety guarantees}{
We present a sketch of the proof of the safety guarantees of \codename{} by
providing the key lemmas.
The detailed proofs of the lemmas are available in Section~\ref{appx:proofs} of the Appendix.

First, we show that well-typedness implies safety for terms.

\begin{lemma}[Safety of terms]
\label{lemma:term-safe}
A well-typed term is safe.
\end{lemma}

Then, by matching the $\ltsend{\exlog}{v}$ and $\ltrecv{\exlog^\prime}{v}$ when
obtaining the execution logs of well-typed compositions, we prove that
well-typedness implies safety also for compositions.

\begin{lemma}[Safety of compositions]
\label{lemma:comp-safe}
A well-typed composition is safe.
\end{lemma}

Then, to account for the looping semantics in programs, we show that well-typedness
for an \codename{} process $\texttt{loop} \{ t \}$
is sufficient to guarantee that any number of $t$s joined together by wait ($\Rightarrow$)
is also well-typed.

\begin{lemma}[Two iterations are sufficient]
\label{lemma:iter}
Let $t$ be an \codename{} term and $t_k, k = 1, 2, \cdots$ be
inductively defined as $t_1 = t$ and $t_{k + 1} = t_k \Rightarrow t$.
If $t_2$ is well-typed, $t_k$ is well-typed for all $k = 2, \cdots$.
\end{lemma}

With the results above, 
the following theorem that describes the main safety guarantees of \codename{}
easily follows.

\begin{theorem}[\codename{} safety guarantees]
\label{thm:main}
A well-typed \codename{} program is safe.
\end{theorem}
}

\section{Proofs}
\label{appx:proofs}

\subsection{Additional Lemmas}

\begin{lemma}
\label{lemma:logtotimestamp}
   If a term $t$ is well-typed and $\tjudgefull{\emptyset}{G}{R}{M}{\emptyset}{e_0}{t : T}$,
   then for every local execution log
   $\exlog = \langle \alpha_0, \cdots, \alpha_k \rangle$ of $t$, there exists
   a timestamp function $\tau_G$ of $G$, such that
   if $\tjudgefull{\Gamma}{G}{R}{M}{C}{e_c}{t^\prime : (e_l, S_d)}$ appears during inference
   of $\tjudgefull{\emptyset}{G}{R}{M}{\emptyset}{e_0}{t : T}$, and
   $\ejudge{\Gamma^\prime}{I^\prime}{M}{\evalto{t^\prime}{\exlog^\prime}{v}{}}$
   appears during inference of
   $\ejudge{\emptyset}{I}{M}{\evalto{t}{\exlog}{v_0}{}}$,
   let $\exlog^\prime = \langle \alpha^\prime_0, \cdots, \alpha^\prime_l \rangle$, then
   $\forall 0 \leq i \leq l : \alpha^\prime_i \subseteq \alpha_{i + \timestamp{G}{e_c}}$
   and $\timestamp{G}{e_c} + l = \timestamp{G}{e_l}$.
   And for all $r \in D, \regdep{\exlog}{v}{D}$,
   there exists $(e, S) \in R(r)$, such that
   $e \leq_G e_l$ and $S_d \leq_G S$.
\end{lemma}
\begin{proof}
    We first show that such a function $\tau_G$, if it exists, is
    a timestamp function of $G$.
    Consider the sub-terms $t^\prime$ that appear
    both in typing inference and evaluation.
   If $\tjudgefull{\Gamma}{G}{R}{M}{C}{e_c}{t^\prime : (e_l, S_d)}$ appears during inference
   of $\tjudgefull{\emptyset}{G}{R}{M}{\emptyset}{e_0}{t : T}$, and
   $\ejudge{\Gamma^\prime}{I^\prime}{M}{\evalto{t^\prime}{\exlog^\prime}{v}{}}$
   appears during inference of
   $\ejudge{\emptyset}{I}{M}{\evalto{t}{\exlog}{v_0}{}}$,
   we show that $\timestamp{G}{e_c} + l = \timestamp{G}{e_l}$ is
   consistent with the timestamp function definition.
   In addition, we show $\forall (x : (k, v^\prime)) \in \Gamma^\prime : \Gamma (x) = (e^\prime_l, S^\prime_d) \to k = \max (0, \timestamp{G}{e^\prime_l} - \timestamp{G}{e_c})$.
    This is shown by
    considering all possibilities
    for the rules applied and for each case replacing one constraint for the timestamp
    with a stricter equation. For example:
    \begin{itemize}
        \item \textsc{T-Cycle} and \textsc{E-Cycle}: $G(e_c) = \graphpred{\#k}{\{e_l\}}, l = k$.
        \item \textsc{T-Wait} and \textsc{E-Wait}: $
        \timestamp{G}{e_c} + l_1 = \timestamp{G}{e^\prime_l},
        \timestamp{G}{e^\prime_l} + l_2 = \timestamp{G}{e_l}, l = l_1 + l_2$.
        \item \textsc{T-Ref} and \textsc{E-Ref}: $
        l = k, G(e_l) = \graphpred{\#0}{\{e_c, e^\prime_l\}}$.
    \end{itemize}

    Let $k$ be the number of all such sub-terms, then there are
    $k$ linear equations, and each equation involves at least one unique
    variable. Hence any subset of those equations contain at least as many
    variables as equations. Therefore, the system of linear equations has at least one
    solution. In other words, $\tau_G$ exists and is a timestamp function of $G$.

    Now we prove that with such a $\tau_G$, $\forall 0 \leq i \leq l : \alpha^\prime_i \subseteq
    \alpha_{i + \tau_G(e_c)}$, where
    $\exlog^\prime = \langle \alpha^\prime_0, \cdots, \alpha^\prime_l \rangle$. This is shown by induction.

    By induction, we can prove that
   for all $r \in D, \regdep{\exlog}{v}{D}$,
   there exists $(e, S) \in R(r)$, such that
   $e \leq_G e_l$ and $S_d \leq_G S$.
\end{proof}

\subsection{Lemma~\ref{lemma:term-safe}}
\begin{proof}
Let $t$ be a well-typed \codename{} term.
From the definition of well-typedness,
$\tjudgefull{\emptyset}{G}{R}{M}{\emptyset}{e_0}{t : T}$.
We show that
for every local execution log $\exlog = \langle \alpha_0, \cdots, \alpha_k \rangle$,
$\ejudge{\emptyset}{I}{M}{\evalto{t}{\exlog}{v_0}{}}$,
the timestamp function in Lemma~\ref{lemma:logtotimestamp}
satisfies that
for every value $v$, if
$\ejudge{\Gamma^\prime}{I^\prime}{M}{\evalto{t^\prime}{\exlog^\prime}{v}{}}$
appears during inference of
$\ejudge{\emptyset}{I}{M}{\evalto{t}{\exlog}{v_0}{}}$,
and
$\tjudgefull{\Gamma}{G}{R}{M}{C}{e_c}{t^\prime : (e_l, S_d)}$ appears
in during inference of
$\tjudgefull{\emptyset}{G}{R}{M}{\emptyset}{e_0}{t : T}$,
let
$a = \timestamp{G}{e_l},
b = \timestamp{G}{\min_{\timepattern{e}{p} \in S_d},
\timestamp{G}{\timepattern{e}{p}}}$,
then $\useset{\exlog}{v} \subseteq [a, b]$ and
for all $D$ such that $\regdep{\exlog}{v}{D}$,
$\mutset{\exlog}{D} \cap [a, b) = \emptyset$.

Consider each member $i \in \useset{\exlog}{v}$. By induction,
it is obvious that one of the following must hold:
\begin{itemize}
    \item $\valueuse{v} \in \alpha^\prime_0$ by
    \textsc{E-IfThen}, \textsc{E-IfElse}, and \textsc{E-RegAssign}. By Lemma~\ref{lemma:logtotimestamp},
    $i = \timestamp{G}{e_c}$
    \item $\valuecreate{v}{S_r}{S_v} \in \alpha^\prime_0$
    by \textsc{E-RegVal}. Similarly, $i = \timestamp{G}{e_l}$
    \item $\valuecreate{v}{S_r}{S_v} \in \alpha^\prime_k$
    by \textsc{E-Cycle}. In this case, $i = \timestamp{G}{e_l}$
\end{itemize}
In each case, we get $i \in [a, b]$.
Thus $\useset{\exlog}{v} \subseteq [a, b]$.

Now we prove for $\regdep{\exlog}{v}{D}, \mutset{\exlog}{D} \cap [a, b) = \emptyset$.
Consider each $i \in \mutset{\exlog}{D}$. By definition,
we have some $r \in D, \regmutate{r} \in \alpha_i$.
By Lemma~\ref{lemma:logtotimestamp}, there must be applications of \textsc{E-RegAssign} and \textsc{T-RegAssign}
where $\timestamp{G}{e_c} = i$ and there exists
$(e, S) \in R(r)$ such that $e \leq_G e_l$ and $S_d \leq_G S$.
Either $e_c <_G e$ or $S \leq_G e_c$.
If $e_c <_G e$, by definition of $<_G$ and $\leq_G$,
we have $i = \timestamp{G}{e_c} < \timestamp{G}{e} \leq_G \timestamp{G}{e_l} = a$. Hence, $i \not\in [a, b)$.
If $S \leq_G e_c$, similarly, we have
$b = \timestamp{G}{S_d} \leq_G \timestamp{G}{S} \leq_G \timestamp{G}{e_c} = i$. Hence, we also have $i \not\in [a, b)$.
Therefore, $\mutset{\exlog}{v} \cap [a, b) = \emptyset$.

By definition of safety, $t$ is safe.
\end{proof}

\subsection{Lemma~\ref{lemma:comp-safe}}

\begin{proof}
Let $\exlog$ be an execution log of $t_1 \parallel_\Sigma t_2$.
By definition, $\exlog$ can be obtained by combining $\exlog_1$
and $\exlog_2$, each an execution log of $t_1$ and $t_2$, respectively.
Since $t_1$ and $t_2$ are well-typed, $t_1$ and $t_2$ are safe, and
$\exlog_1, \exlog_2$ are also safe.
By definition of safety,
for every value $v$, there exists $a_1, b_1, a_2, b_2$, such that
$\useset{\exlog_1}{v} \cup \ltsend{\exlog_1}{v} \subseteq [a_1, b_1] \subseteq \ltrecv{\exlog_1}{v},
\regdep{\exlog}{v}{D_1},
\mutset{\exlog_1}{D_1} \cap [a_1, b_1) = \emptyset$, and
$\useset{\exlog_2}{v} \cup \ltsend{\exlog_2}{v} \subseteq [a_2, b_2] \subseteq \ltrecv{\exlog_2}{v},
\regdep{\exlog}{v}{D_2},
\mutset{\exlog_2\\}{D_2} \cap [a_2, b_2) = \emptyset$.

For $i \in \{1, 2\}$,
if a $\valuecreate{v}{S_r}{S_v}$ appears in $\exlog_i$,
or, if no $\valuecreate{v}{S_r}{S_v}$ appears in either $\exlog_i$ or
$\exlog_{3 - i}$ but $\ltrecv{\pi.m}{v}$ appears in $\exlog_i$,
we say that $\exlog_i$ owns $v$.
Obviously every $v$ that appears in $\exlog$ is owned by
either $\exlog_1$ or $\exlog_2$ but not both.
We show that the following $a, b$ satisfies that
$\useset{\exlog}{v} \cup \ltsend{\exlog}{v} \subseteq [a, b] \subseteq \ltrecv{\exlog}{v},
\regdep{\exlog}{v}{D},
\mutset{\exlog}{D} \cap [a, b) = \emptyset$:
\begin{enumerate}
    \item If $v$ does not appear in $\exlog$, then $a = a_1, b = b_1$.
    \item If $v$ appears in $\exlog$, and is owned by $\exlog_i$,
    $a = a_i, b = b_i$.
\end{enumerate}

Case~1 is trivial.

For Case~2, by induction on the structure of $\depset{\exlog}{v}$,
it is easy to obtain that $\useset{\exlog}{v} \cup \ltsend{\exlog}{v}
\subseteq \useset{\exlog_i}{v} \cup \ltsend{\exlog_i}{v}$ and
$\ltrecv{\exlog_i}{v} \subseteq \ltrecv{\exlog}{v}$.
Therefore, we get $\useset{\exlog}{v} \cup \ltsend{\exlog}{v}
\subseteq [a_i, b_i] \subseteq \ltrecv{\exlog}{v}$.
Now we prove that $\mutset{\exlog}{D} \cap [a_i, b_i) = \emptyset$.
Without loss of generality, we assume $i = 1$.

We use induction on $\depset{\exlog}{v}$.
Consider the following cases:
\begin{enumerate}
    \item $\depset{\exlog}{v} = \emptyset$. In this case,
     either
    $\valuecreate{v}{S_r}{\emptyset}$ or $\ltrecv{\pi.m}{v}$ appears
    in both $\exlog_1$ and $\exlog$.
    In both cases, $\mutset{\exlog}{D} = \mutset{\exlog_1}{D_1}$.
    Since $\mutset{\exlog_1}{D_1} \cap [a_1, b_1) = \emptyset$,
    $\mutset{\exlog}{D} \cap [a_1, b_1) = \emptyset$.
    \item $\depset{\exlog}{v} = S_v$. In this case,
    $\valuecreate{v}{S_r}{S_v}$ is in both $\exlog_1$ and $\exlog$.
    Consider each $u \in S_v$.
    Either $u$ is owned by $\exlog_1$, or
    it is owned by $\exlog_2$.
    Let $a^\prime, b^\prime$ be selected such that
    $\useset{\exlog_j}{u} \cup \ltsend{\exlog_j}{u}
    \subseteq [a^\prime, b^\prime] \subseteq \ltrecv{\exlog_j}{u}$
    and $\mutset{\exlog_j}{D^\prime} \cap [a^\prime, b^\prime)$, where $\exlog_j$ is the owner of $u$.
    Let $a_0, b_0$ be selected such that
    $\useset{\exlog_1}{u} \cup \ltsend{\exlog_1}{u}
    \subseteq [a_0, b_0] \subseteq \ltrecv{\exlog_1}{u}$
    and $\mutset{\exlog_1}{D_0} \cap [a_0, b_0)$.
    If $j = 1$, then $a_0 = a^\prime_u, b_0 = b^\prime_u$.
    If $j = 2$, there must be a send operation involving $u$ in $\exlog_2$
    and a matching receive operation in $\exlog_1$.
    We have $[a_0, b_0] \subseteq \ltrecv{\exlog_1}{u} \subseteq
    \ltsend{\exlog_2}{u} \subseteq [a^\prime_u, b^\prime_u]$.
    In both cases, we have $[a_0, b_0] \subseteq [a^\prime_u, b^\prime_u]$.
    By induction assumptions, $[a^\prime, b^\prime) \cap \mutset{\exlog}{D_u} =
    \emptyset$, hence
    $[a_0, b_0) \cap \mutset{\exlog}{D_u} = \emptyset$.
    Combining all $u \in S_v$, by definition of $\ltrecv{\exlog_1}{v}$, $\mutset{\exlog_1}{v}$, and
    $\mutset{\exlog}{v}$:
    $[a, b] \subseteq \bigcap_{u \in S_v}\ltrecv{\exlog_1}{u}
    \subseteq \bigcap_{u \in S_v}[a^\prime_u, b^\prime_u],
    \mutset{\exlog}{v} = \mutset{\\\exlog_1}{v} \cup \bigcup_{u \in S_v}\mutset{\exlog}{u}$.
    Hence $[a, b) \subseteq \bigcap_{u \in S_v}[a^\prime_u, b^\prime_u)$,
    and $[a, b) \cap \mutset{\exlog}{v} \subseteq
    \bigcap_{u \in S_v}[a^\prime_u, b^\prime_u) \cap
    \bigcup_{u \in S_v}\mutset{\exlog}{u} = \emptyset$.
\end{enumerate}

By induction, if $v$ is owned by $\exlog_i$,
$\useset{\exlog}{v} \cup \ltsend{\exlog}{v} \subseteq [a_i, b_i] \subseteq \ltrecv{\exlog}{v}$ and
$[a_i, b_i) \cap \mutset{\exlog}{D} = \emptyset$.
Combining Case~1 and Case~2, we have shown that
for all $v$,
there exist such $a$ and $b$.
Therefore, the composition $t_1 \parallel_\Sigma t_2$ is safe.

\end{proof}

\subsection{Lemma~\ref{lemma:iter}}
\begin{proof}
We show that for $k \geq 2$, if $t_k$ is well-typed, $t_{k + 1}$ is also well-typed.
By induction, this implies that if $t_2$ is well-typed, $t_k (k = 2, \cdots)$ are all
well-typed.

Since $t_k$ is well-typed, we have $\tjudgefull{\emptyset}{G}{R}{M}{\emptyset}{e_0}{t_k : T}$.
Because $t_k = t_{k - 1} \texttt{ => } t$, there exists
$\tjudgefull{\emptyset}{G}{R}{M}{C_1}{e_0}{t_{k - 1} : (e_1, S_1)}$
and
$\tjudgefull{\emptyset}{G}{R}{M}{C_2}{e_1}{t : (e_2, S_2)}$ which
appear during inference.
It is obvious that $e_1$ is a cut vertex in $G$, i.e.,
there exists a partition of $V = V_1 \cup V_2 \cup \{e_1\}$, such that
all paths between $V_1$ and $V_2$ go through $e_1$, and
it can be found such that $V_2 \cup \{e_1\}$ is the set of
all nodes that appear in the inference rules used to obtain
$\tjudgefull{\emptyset}{G}{R}{M}{C_2}{e_1}{t : (e_2, S_2)}$.
Let $G_2$ be the subgraph of $G$ with $V^\prime = V_2 \cup \{e_1\}$.
Let $G^\prime_2$ be a graph obtained by relabelling nodes of $G_2$
such that $e_1$ is relabelled $e_2$ and nodes in $V_2$ are relabelled
to nodes in $V_3$, where $V_3 \cap V = \emptyset$.
Now let $G^\prime = G \cup G^\prime_2$.
Obviously, assuming $<_G$ and $\leq_G$ always hold,
we can obtain
$\tjudgefull{\emptyset}{G^\prime}{R^\prime}{M}{C^\prime}{e_0}{t_{k + 1} : T^\prime}$
such that
the same nodes appear in rules inferring for $t_k$, and additionally there are
rules inferring for $t$ that simply map nodes used inferring
$\tjudgefull{\emptyset}{G}{R}{M}{C_2}{e_1}{t : (e_2, S_2)}$ from $V_2 \cup \{e_1\}$ to
$V_3 \cup \{v_2\}$.
Therefore, if $t_{k + 1}$ is not well-typed, there must be some unattainable $<_G$
or $\leq_G$ that appear in those rules.
Consider different cases:
\begin{itemize}
    \item Some $e_a <_G e_b$ or $e_a \leq_G e_b$, which only involves nodes but not event patterns,
    does not hold.
    Obviously, $\{e_a, e_b\} \in V_1 \cup \{e_1\}$ or
    $\{e_a, e_b\} \in V_2 \cup \{e_1\}$ or $\{e_a, e_b\} \in V_3 \cup \{e_2\}$.
    This always implies that a corresponding typing judgment does not hold for inferring
    well-typedness of $t_k$, contradicting the assumption.
    \item Some typing judgment that involves $\timepattern{e_a}{p}$ does not hold.
    This similarly imply a contradiction a rule involved in inferring the well-typedness
    of $t_k$ does not hold.
\end{itemize}

By contradiction, $t_{k + 1}$ is well-typed.

\end{proof}

\end{document}